\documentclass[aps,prl,twocolumn,superscriptaddress,showpacs]{revtex4-1}
\usepackage[utf8]{inputenc}
\usepackage[pdftex]{epsfig}
\usepackage{dcolumn}
\usepackage{graphicx}
\usepackage{latexsym}
\usepackage{amsmath}
\usepackage[pdftex,colorlinks]{hyperref}
\usepackage{rotating}
\usepackage{braket}
\usepackage{relsize}
\hypersetup{colorlinks,%
                 citecolor=blue,%
                 filecolor=blue,%
                 linkcolor=blue,%
                 urlcolor=blue,%
                 pdftex}
\usepackage{color}
\usepackage[dvipsnames]{xcolor}
\usepackage{array,booktabs}

\begin{document}
\title{Intermultiplet transitions and magnetic long-range order in Sm-based pyrochlores}

\author{Viviane Pe\c canha-Antonio}
\email{v.pecanha.antonio@fz-juelich.de}
\affiliation{J\"ulich Centre for Neutron Science (JCNS) at Heinz Maier-Leibnitz Zentrum (MLZ), Forschungszentrum J\"ulich GmbH, Lichtenbergstr. 1, 85748 Garching, Germany}
\affiliation{Physik-Department, Technische Universit\"at M\"unchen, D-85747 Garching, Germany}

\author{Erxi Feng}
\affiliation{J\"ulich Centre for Neutron Science (JCNS) at Heinz Maier-Leibnitz Zentrum (MLZ), Forschungszentrum J\"ulich GmbH, Lichtenbergstr. 1, 85748 Garching, Germany}

\author{Xiao Sun}
\affiliation{J\"ulich Centre for Neutron Science (JCNS) and Peter Gr\"unberg Institut (PGI), Forschungszentrum J\"ulich GmbH, D-52425 J\"ulich, Germany}

\author{Devashibhai Adroja}
\affiliation{ISIS Facility, Rutherford Appleton Laboratory, Chilton, Didcot OX11 0QX, United Kingdom}

\author{Helen C. Walker}
\affiliation{ISIS Facility, Rutherford Appleton Laboratory, Chilton, Didcot OX11 0QX, United Kingdom}

\author{Alexandra Gibbs}
\affiliation{ISIS Facility, Rutherford Appleton Laboratory, Chilton, Didcot OX11 0QX, United Kingdom}

\author{Fabio Orlandi}
\affiliation{ISIS Facility, Rutherford Appleton Laboratory, Chilton, Didcot OX11 0QX, United Kingdom}

\author{Yixi Su}
\email{y.su@fz-juelich.de}
\affiliation{J\"ulich Centre for Neutron Science (JCNS) at Heinz Maier-Leibnitz Zentrum (MLZ), Forschungszentrum J\"ulich GmbH, Lichtenbergstr. 1, 85748 Garching, Germany}

\author{Thomas Br\"uckel}
\affiliation{J\"ulich Centre for Neutron Science (JCNS) and Peter Gr\"unberg Institut (PGI), Forschungszentrum J\"ulich GmbH, D-52425 J\"ulich, Germany}


\begin{abstract}

We present bulk and neutron scattering measurements performed on the isotopically enriched $^{154}\mathrm{Sm_2Ti_2O_7}$ and $^{154}\mathrm{Sm_2Sn_2O_7}$ samples. Both compounds display sharp heat capacity anomalies, at 350 mK and 440 mK, respectively. Inelastic neutron scattering measurements are employed to determine the crystalline electric field (CEF) level scheme, which includes transitions between the ground-state and first excited $J$ multiplets of the $\mathrm{Sm}^{3+}$ ion. To further validate those results, the single-ion magnetic susceptibility of the compounds is calculated and compared with the experimental DC-susceptibility measured in low applied magnetic fields. It is demonstrated that the inclusion of intermultiplet transitions in the CEF analysis is fundamental to the understanding of the intermediate and, more importantly, low temperature magnetic behaviour of the Sm-based pyrochlores. Finally, the heat capacity anomaly is shown to correspond to the onset of an all-in-all-out long-range order in the stannate sample, while in the titanate a dipolar long-range order can be only indirectly inferred.

\end{abstract}

\maketitle

\section{Introduction}

Pyrochlores, compounds of chemical formula $\mathrm{A_2B_2O_7}$, where A is a rare-earth ion and B is a transition metal, present an essential requirement for a geometrically frustrated material: the crystallographic lattice of corner-sharing tetrahedra forming in the cubic $Fd\bar{3}m$ space group. The frustration, which arises as a direct consequence of the lattice geometry, is the mechanism that, in theory, may inhibit long-range order at temperatures as low as absolute zero \cite{Balents}. Paradoxically, pyrochlores are likewise interesting due to their magnetic ground-states, which in general, and despite the frustration, present magnetic long-range order.

One of the primordial investigations conducted in rare-earth magnets is the search for crystal electric field (CEF) excitations. The free rare-earth ion is subject to strong spin-orbit coupling, which turns the total angular momentum $J=L+S$ into a good quantum number \cite{abragam}. When in the pyrochlore lattice, the full-rotation symmetry of the free-ion is reduced to the point group $D_{3d}$. In this geometry, the trigonal charge environment around the rare-earth splits each of the $(2J+1)$-fold degenerated multiplets into a series of singlets or doublets. As $\mathrm{Sm^{3+}}$ possesses an odd number of electrons in its unfilled $4f$ orbital, all possible $J$'s will be half-integers and each multiplet can be shown to split into $(J+1/2)$ \emph{Kramers-doublets} \cite{Hutchings}. The crystal field plays a pivotal role in the compounds ground-state magnetic anisotropy, and consequently, in the ultimate magnetic ordering that may be assumed by the system at low-temperatures. 

From the material syntheses point of view, pyrochlores also present interesting properties. The family of the titanates, for which the B atom is a Ti$^{4+}$, and of the stannates, where B is a Sn$^{4+}$, are possibly the two most studied \cite{RevModPhys.82.53}. Under standard solid state synthesis conditions, the pyrochlore phase formation, with few exceptions \cite{RevModPhys.82.53}, obeys a strict rule. If the ionic radius ratio $R_{\mathrm{A^{3+}}}/R_{\mathrm{B^{4+}}}$ assumes values between 1.46 and 1.78, the cubic, ordered $\mathrm{A_2B_2O_7}$ phase will form and remain stable at 1 atm pressure \cite{SUBRAMANIAN198355}. The stannate family is the only one that satisfies this condition for all the rare-earth ions, from Lu$^{3+}$ to La$^{3+}$. The titanates family is more restricted, and the ratio $R_{\mathrm{Sm^{3+}}}/R_{\mathrm{Ti^{4+}}}=1.78$ lies at the border of the stability-field diagram.

This work is based on two, little investigated members of those families: the Sm-based titanate $\mathrm{Sm_2Ti_2O_7}$ and stannate $\mathrm{Sm_2Sn_2O_7}$. The former compound is well known for its photocatalytic \cite{S0925838817304541}, nuclear waste storage \cite{S0955221914003628} and electronic applications \cite{S0924013699002113}. Three key studies on magnetic frustration have been conducted on the titanate \cite{PhysRevB.77.054408,Malkin,PhysRevB.98.100401}, and are going to be cited recurrently here. For the stannate, reported work is limited to static magnetic susceptibility results, which were presented by Bondah-Jagalu \emph{et al.} \cite{nrc_cjp79_1381}

Singh \emph{et al.} \cite{PhysRevB.77.054408} shows results of DC-susceptibility ($\chi_\mathrm{dc}$), heat capacity ($\mathrm{C_p}$) and Raman spectroscopy measurements conducted on a single-crystal sample of $\mathrm{Sm_2Ti_2O_7}$. The analysis of $\chi_\mathrm{dc}$ and $\mathrm{C_p}$ demonstrated that the dipolar and exchange interactions in the titanate have smaller energy scales when compared to other members of the pyrochlore family. Down to 2 K, no signal of spin-freezing could be detected. A maximum in the susceptibility, at $\sim140$ K, was interpreted as a consequence of the single-ion properties of the Sm$^{3+}$ ion. Raman spectroscopy indicated the presence of four active low-energy modes attributed to CEF excitations, while the ground-state $J=5/2$ of Sm$^{3+}$ is expected to split into a maximum of three doublets. Malkin \emph{et al.} \cite{Malkin} reanalyses the susceptibility data measured by Singh \emph{et al.} in order to estimate the CEF parameters of $\mathrm{Sm_2Ti_2O_7}$. Despite the excellent agreement between experimental data and fitting, the CEF levels deduced in Malkin \emph{et al.} depart strongly from the ones obtained in the Raman spectroscopy study of Ref. \cite{PhysRevB.77.054408}.

Recently, Mauws \emph{et al.} \cite{PhysRevB.98.100401} presented a more complete set of experiments on another single crystal sample of $\mathrm{Sm_2Ti_2O_7}$, this time enriched with the isotope $^{154}$Sm. It is shown that the titanate displays a heat capacity anomaly at $T^{Ti}_\mathrm{N}=350\ \mathrm{mK}$, associated with the development of an antiferromagnetic all-in-all-out long-range order in the sample. Additionally, inelastic neutron scattering is employed to determine a third set of crystal electric field levels, which is partially inconsistent with both \cite{PhysRevB.77.054408} and \cite{Malkin}. 

On the one hand, the information contained in the structure factor $S(|\mathbf Q |,\omega)$ measured using inelastic neutron scattering could provide an explanation to the excessive number of CEF levels reported by Singh \emph{et al.} \cite{PhysRevB.77.054408} and their diverging position in energy with the calculated levels of Malkin \emph{et al.} \cite{Malkin}. On the other, two challenges related with the research of Sm-pyrochlores using neutron scattering techniques must be considered. Firstly, in its natural abundance, Sm is a strong neutron absorber. Moreover, the ion has the smallest ordered magnetic moment of all the trivalent magnetic rare-earths. The first problem can be circumvented by enriching the samples with $^{154}$Sm, which is the most abundant, less absorbing isotope of the element. The second problem is not directly solvable and, for this reason, we combine here the studies of two compounds. Our goal is to reach common conclusions that could be established by the systematic analysis and comparison of these two slightly different pyrochlores.

In this work, it is shown that not only the titanate, at $T^{Ti}_\mathrm{N}=350\ \mathrm{mK}$, but also the stannate, at $T^{Sn}_\mathrm{N}=440\ \mathrm{mK}$, presents a sharp anomaly in $\mathrm{C_p}$. Using inelastic neutron scattering, the crystal electric field transitions between the ground-state doublet and the levels of the first excited multiplet of the Sm$^{3+}$ ion are measured and fit on both compounds, in order to solve the CEF scheme of the ion in the pyrochlore lattice. Our analyses demonstrate that the values of ground-state magnetic moment, when intermultiplet transitions are not neglected, are strongly suppressed from those predicted when considering uniquely the ground-state CEF splitting. We extend our analysis calculating the $\chi_{CEF}$ of the titanate and stannate and comparing it with the $\chi_\mathrm{dc}$ measured in a low applied magnetic field. Finally, the possible magnetic long-range order developing in the systems below the phase transition temperature is investigated.

\section{Experimental details}

The samples were synthesised via the solid state reaction method at the J\"ulich Centre for Neutron Science (JCNS) at the Heinz Maier-Leibnitz Zentrum (MLZ), in Garching, Germany. After initial drying, stoichiometric quantities of 99.999\% pure $\mathrm{Sm_2O_3}$, or the nominally 98.5\% isotopically enriched $\mathrm{^{154}Sm_2O_3}$, were mixed with the transition metal oxides $\mathrm{TiO_2}$ and $\mathrm{SnO_2}$ and sintered at 1200 $^{\circ}$C and 1300 $^{\circ}$C, respectively, in four rounds of 24 hours with three intermediate grindings. To check the sample quality, neutron diffraction was performed at the instrument HRPD at ISIS. Around $2\ \mathrm{g}$ of sample were loaded in 3 mm diameter, vanadium cylindrical sample cans. The samples were cooled in a liquid He cryostat down to $5\ \textrm{K}$ and data were collected at 5, 20 and $100\ \mathrm{K}$. Apart from those, room temperature diffraction patterns (labeled as $300\ \mathrm{K}$ below) were measured without the use of any sample environment. This procedure reduces the background considerably.  

Static susceptibility measurements were performed on our isotopically enriched samples down to 5 K at a Quantum Design MPMS SQUID magnetometer. Around 20 mg of each powder sample was pressed and loaded in a polymeric container. The diamagnetic signal of this container was subtracted from the raw SQUID voltage and a dipolar response function was fitted to the corrected data for each temperature point measured. Heat capacity at constant pressure was measured at temperatures ranging from 100 mK up to 4 K using $\sim0.3\ \mathrm{mg}$ sintered pellets of the $\mathrm{^{154}Sm_2Ti_2O_7}$ and $\mathrm{Sm_2Sn_2O_7}$ samples at a Quantum Design PPMS equipped with a dilution insert.

Neutron scattering data were collected exclusively for the $\mathrm{^{154}Sm_2Ti_2O_7}$ and $\mathrm{^{154}Sm_2Sn_2O_7}$ samples. Inelastic neutron scattering measurements were performed at the time-of-flight (TOF) spectrometer MERLIN at ISIS \cite{merlin}. Approximately 0.7 g of the titanate and 1.4 g of the stannate were packed in aluminium foil, subsequently curled up to form hollow cylinders of $\sim2$ cm diameter and height. The samples were sealed in aluminium cans and cooled down to $5\ \mathrm{K}$ in a close cycle refrigerator (CCR) cryostat in He exchange gas. Two experimental configurations were employed. First, the instrument gadolinium chopper was operated in repetition-rate multiplication mode, which enables the simultaneous measurement of incident energies $E_i$ of 10, 19 and, with higher flux, 50 meV for a rotation frequency of 250 Hz. Similarly, for a frequency of 400 Hz, we collected data for $E_i=17,\ 28,\ 54\ \textup{and}\ 150\ \textup{meV}$. As we proceeded to measurements with higher incident energies, the Gd chopper was substituted by a sloppy chopper and data for a single $E_i=300\ \textup{meV}$ at 450 Hz were collected.

Unpolarised neutron diffraction measurements were carried out at very low temperatures at the instruments DNS at MLZ, and WISH at ISIS \cite{wish}. Only the $\mathrm{^{154}Sm_2Ti_2O_7}$ was measured at WISH. The powder was loaded in a copper cylinder can sealed in He atmosphere. Sub-Kelvin temperatures were achieved with help of an Oxford dilution insert placed in an Oxford cryostat. Measurements were performed at 50 mK, 700 mK and 10 K for approximately 6 hours at each temperature. At DNS, data were collected for both $\mathrm{^{154}Sm_2Ti_2O_7}$ and $\mathrm{^{154}Sm_2Sn_2O_7}$ using a FRM II standard cryogen-free dilution cryostat. The powders were loaded in an annular cooper cylinder can and sealed in He atmosphere. To speed up the cooling process, $\sim 0.1$ ml of deuterated ethanol was added to the samples measured at DNS. This procedure was already used by us before and resulted in the successful cooling of powder samples in the dilution temperature regime. For the stannate, data were collected for 12 hours at 75 and 600 mK and for the titanate for at least 18 hours at 170 and 600 mK. 

\section{Results}

\subsection{Sample stoichiometry}
We performed simultaneous Rietveld refinements on the data collected at the three different detector banks of HRPD using the software GSAS. The crystallographic model employed in the refinement contains the coordinates and occupancies corresponding to the second origin choice of the ${F d \bar{3} m}$ space group. In order to have a better quantitative estimation of the degree of enrichment in our sample, we refine a double occupancy at the rare-earth site. This position is chosen, initially, to be occupied 98.5\% by $^{154}$Sm ions, following the raw material fabricant specifications. The remaining 1.5\% sites are supposed to be filled with the other natural isotopes of the atom. The scattering length of these 1.5\% sites is calculated using the individual scattering lengths given in Ref. \cite{neutron_sclengths}, normalised by the respective isotopic natural abundance. 

We note that some uncertainty concerning the tabulated values of the scattering length for $^{154}$Sm has been reported in the literature \cite{kennedy_sclengths}. The precise knowledge of this value would be essential for an accurate determination of the site occupancies. To systematically analyse and compare our data, we use in the refinements the tabulated values of Ref. \cite{neutron_sclengths}, which is historically the reference for neutron scattering lengths. 

\begin{figure}
\includegraphics[trim={0cm 1.5cm 0cm 5.cm},clip,width=7.9cm,angle=270]{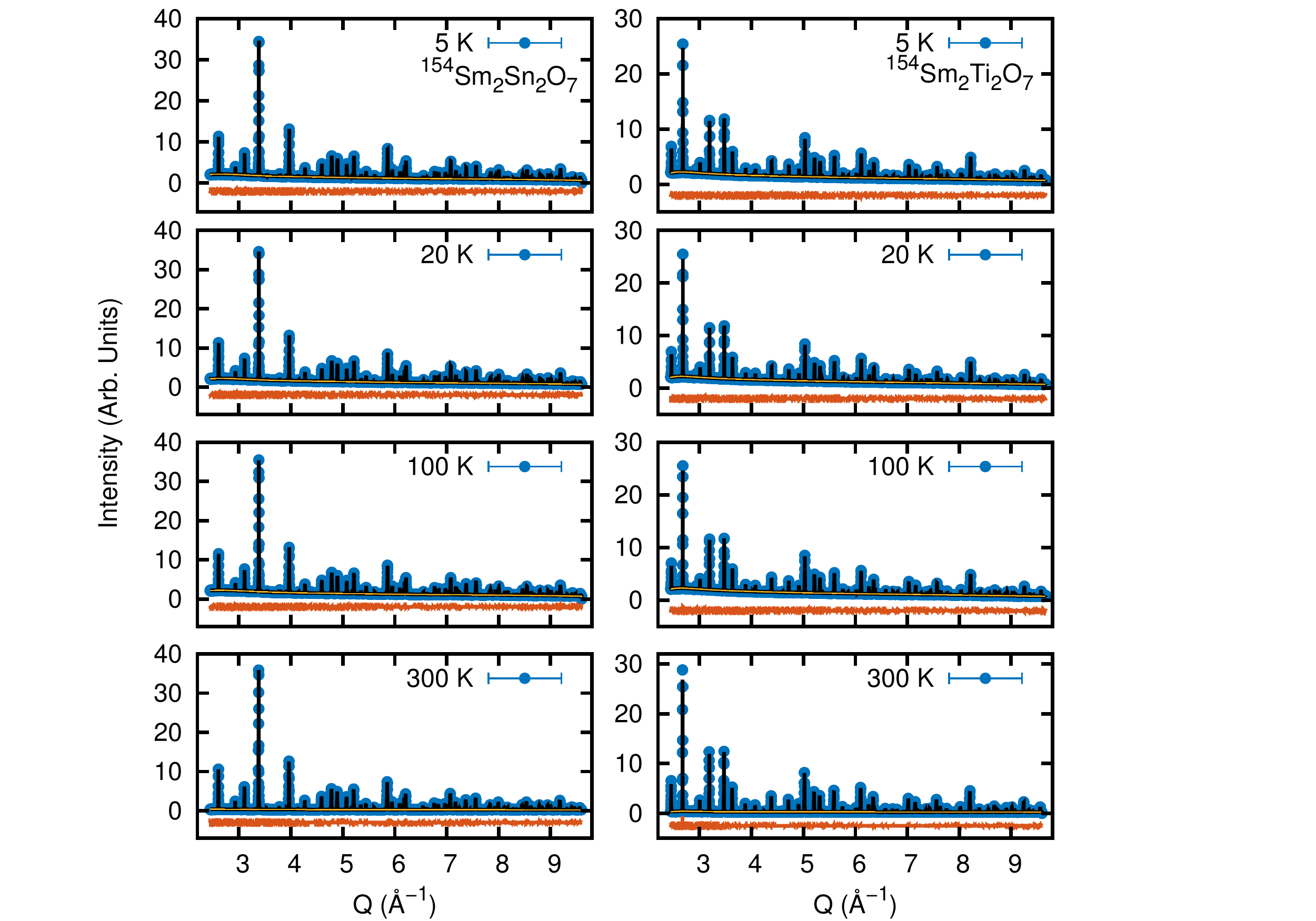}
\caption{Neutron diffraction of stannate (left) and titanate (right) measured at the highest resolution bank of HRPD (blue circles), from top to bottom in order of increasing temperature. The black continuos lines show the best refinement, while the red line below each plot shows the difference between experimental and refined profiles.}\label{SmTO_1_2_alltemp}
\end{figure}

The refined lattice parameters, oxygen $48f$ free position $x$ and fitting $R_{wp}$ are summarised in Table \ref{tab:tab4}. The data collected for stannate and titanate at the highest resolution detector bank of HRPD is shown in Fig. \ref{SmTO_1_2_alltemp}. For all the measured temperatures, the relative occupancy at the $16d$ position was allowed to vary, while the total occupancy of this site was constrained to be identical to one. The final values are consistent with each other and demonstrate that our samples are enriched with $\sim98\%$ of $^{154}\mathrm{Sm}$ isotope. 

\begin{table}
\centering
\tabcolsep=8pt
\renewcommand{\arraystretch}{1.2}
\begin{tabular}{c c c c}
\toprule[1pt]
Temperature & Lattice & \emph{x} &$R_{wp}$\\
  (K)& Parameter (\AA)&&\\
 \cmidrule[0.3pt](l{.75em}r{.75em}){1-4}
 \multicolumn{4}{c }{$\mathrm{^{154}Sm_2Sn_2O_7}$}\\
\cmidrule[0.3pt](l{.75em}r{.75em}){1-4}
5 & 10.49775(2)&0.33379(4) &0.0224\\
20 & 10.49779(5) & 0.33378(4)&0.0229\\
100 & 10.49924(1) &0.33371(4)&0.0242\\
300 &10.51140(4) &0.33351(3) &0.0397\\
 \cmidrule[0.3pt](l{.75em}r{.75em}){1-4}
\multicolumn{4}{c }{$\mathrm{^{154}Sm_2Ti_2O_7}$}\\
\cmidrule[0.3pt](l{.75em}r{.75em}){1-4}
5 & 10.22182(4)&0.32637(5) &0.0222\\
20 & 10.22224(5) & 0.32640(5)&0.0224 \\
100 & 10.22645(5) &0.32634(5)&0.0248 \\
300 &10.24148(4) &0.32611(4) & 0.0450 \\
\bottomrule[0.5pt]
\end{tabular}
\caption{Best refined parameters for $\mathrm{^{154}Sm_2Sn_2O_7}$ and $\mathrm{^{154}Sm_2Sn_2O_7}$ sample at several temperatures. The refined profiles of the highest resolution detector bank are shown in Fig. \ref{SmTO_1_2_alltemp}. }\label{tab:tab4}
\end{table}

No phase impurity is detected in the neutron scattering data of both samples. The lattice parameter of the stannate is larger than that of the titanate, as expected, due to the larger ionic radius of the $\mathrm{Sn^{3+}}$ \cite{RevModPhys.82.53}. Upon warming from 5 to $20\ \mathrm{K}$, only a very slight increase in the lattice parameter is noted. At room temperature, the lattice parameters and $x$ values are in close agreement with the ones reported in Subramanian \emph{et al.} \cite{SUBRAMANIAN198355}.

\subsection{Heat capacity and magnetic entropy}

\begin{figure}
\includegraphics[trim={5cm 0.3cm 3.5cm 0cm},clip,width=3.2cm,angle=270]{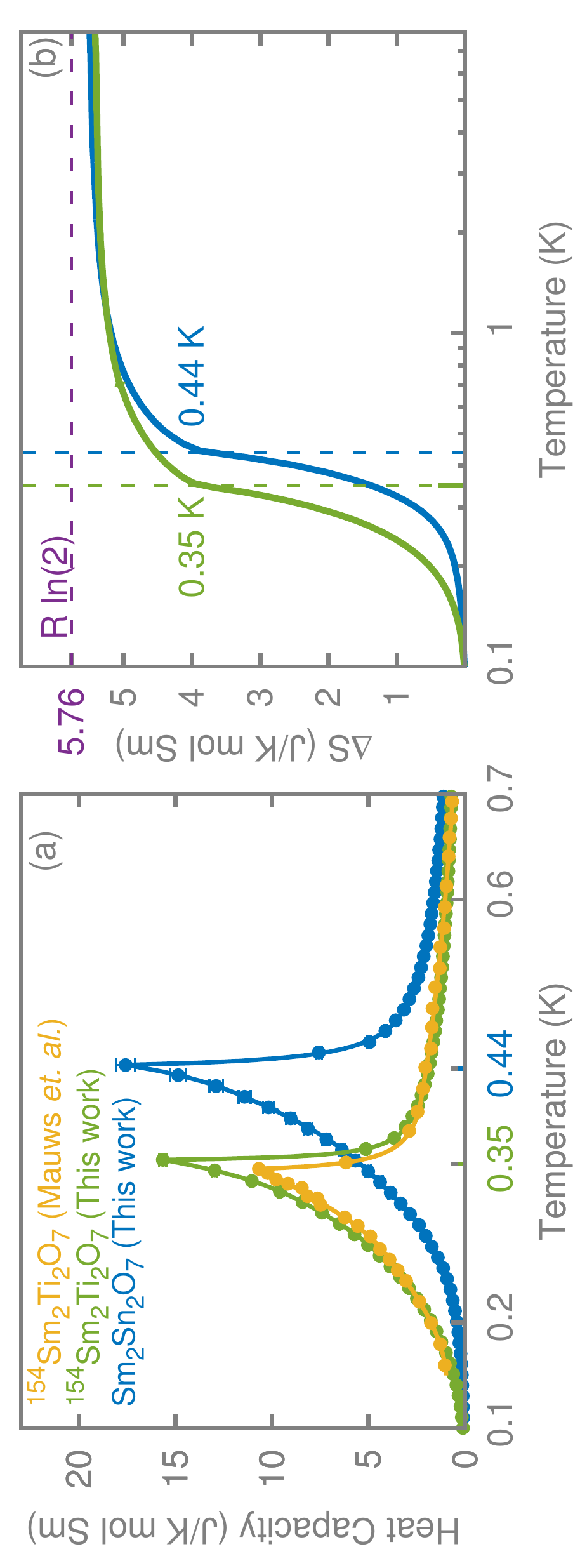}
\caption{(a) Low temperature heat capacity measured for $\mathrm{^{154}Sm_2Ti_2O_7}$ and $\mathrm{Sm_2Sn_2O_7}$ (green and blue filled circles, respectively). For comparison, data of $\mathrm{^{154}Sm_2Ti_2O_7}$ published in Ref. \cite{PhysRevB.98.100401} are also shown (yellow filled circles). The lines are guides to the eye. (b) Estimated magnetic entropy for $T<8\ \mathrm{K}$. The phonon contribution to $\mathrm{C_p}$, within this temperature range, was neglected. The vertical dashed lines highlight the phase transition temperatures of both compounds. The horizontal dashed line shows the value of $R\ \mathrm{ln(2)}$, the magnetic entropy expected from an ordered doublet ground-state.}\label{HC_entropy_multiplot}
\end{figure}

In Fig. \ref{HC_entropy_multiplot}\hyperref[HC_entropy_multiplot]{(a)}, data of heat capacity at constant pressure for $\mathrm{^{154}Sm_2Ti_2O_7}$ and $\mathrm{Sm_2Sn_2O_7}$ are shown. A strong, similar anomaly appears in both samples at temperatures $T^{Ti}_\mathrm{N}=350\ \mathrm{mK}$, for the titanate, and $T^{Sn}_\mathrm{N}=440\ \mathrm{mK}$, for the stannate. Clearly, the same behaviour of $\mathrm{C_p}$ is expected also on the isotopically enriched $\mathrm{^{154}Sm_2Sn_2O_7}$. Along with data of this work, the heat capacity of Mauws \emph{et al.} \cite{PhysRevB.98.100401} is reproduced. The anomaly of the single-crystal sample of Ref. \cite{PhysRevB.98.100401} has a smaller intensity than the anomaly of our powder, but its position and sharpness are very similar.

The similarities between the $\mathrm{C_p}$ anomalies indicate that $\mathrm{^{154}Sm_2Ti_2O_7}$ and $\mathrm{Sm_2Sn_2O_7}$ may present comparable physical behaviour below phase transitions. Interestingly, another Sm-pyrochlore with the B site occupied by a different transition metal ion, the zirconate $\mathrm{Sm_2Zr_2O_7}$, was shown to not display any sharp anomaly down to the lowest measured temperatures ($\sim 0.1$ K) \cite{xu_jianhui}. Instead, the zirconate shows a broad peak in $\mathrm{C_p}$ centred at around 500 mK, which was not associated to the development of long-range order in the sample \cite{xu_jianhui}. 

The magnetic entropy $\Delta S(T)$ is quantified from the heat capacity performing numerically $\int_{T_0}^{T}\frac{C_p(T')}{T'}dT'$, where $T$ is the sample temperature. The result of this integration is shown in Fig. \ref{HC_entropy_multiplot}\hyperref[HC_entropy_multiplot]{(b)}. The phonon contribution to the heat capacity is usually estimated by fitting a Debye $\propto{T^3}$ curve to low temperature data and extrapolating the result to temperatures below 10 K. However, there is some arbitrariness in the interval to which the Debye model can be fitted. We tried several for $10\ \mathrm{K}< T< 20\ \mathrm{K}$, but all returned an overestimated phonon contribution, producing heat capacities higher than the measured ones at lower temperatures. Therefore, we follow the procedure also adopted by Singh \emph{et al.} \cite{PhysRevB.77.054408} and neglect the phononic contribution to the heat capacity below 8 K.

The magnetic entropy at low temperatures nearly reaches asymptotically $R\ \mathrm{ln(2)}$. Thus, we recognise the ground-state of the system as a well-isolated doublet, a characteristic shared between the majority of the pyrochlore compounds (notorious exceptions are, for example, Gd and Eu-pyrochlores \cite{RevModPhys.82.53}). Considering that samarium is a Kramers ion, this doublet ground-state is protected and its degeneracy cannot be lifted by any perturbation that does not break time-reversal symmetry. 

\subsection{Crystal electric field excitations}
In the point charge model, the potential of the crystalline electric field is treated as a perturbation $H_\mathrm{CEF}$ to the free-ion Hamiltonian. The Sm$^{3+}$ sites in pyrochlores have the $D_{3d}$ symmetry and the pertinent $H_\mathrm{CEF}$ is given by \cite{rosenkranz}
\begin{eqnarray}
\label{eq:eq1}
H_\mathrm{CEF}&=B_{0}^{2}C_{0}^{2}+B_{0}^{4}C_{0}^{4}+B_{3}^{4}(C_{-3}^{4}-C_{3}^{4})+B_{0}^{6}C_{0}^{6}\nonumber \\
 &+B_{3}^{6}(C_{-3}^{6}-C_{3}^{6})+B_{6}^{6}(C_{-6}^{6}-C_{6}^{6}), 
\end{eqnarray}
where the $B_{q}^{k}$ are the crystal field parameters and the $C_{q}^{(\pm k)}$ are the Wybourne tensor operators. 


The choice of basis for the matrix representation of $H_\mathrm{CEF}$ is not unique. The most straightforward restriction, which results in the well-known \emph{Stevens' operator formalism}, is to reduce the basis to a single $\ket{L,S,J}$ level \cite{stevens}. This simplification not only limits the CEF analysis to the ground-state $J=5/2$ multiplet but also allows the replacement of the tensor operators $C_{q}^{(\pm k)}$ by the Stevens' operator equivalents. The interchange between the two formalisms requires, additionally, a rescaling of the crystal field parameters \cite{Hutchings}. The parameters obtained in this work and the ones calculated using the basis states on which the Stevens operators act, called here of $S_{k}^{q}$, can be related by    
\begin{equation}\label{eq2}
\begin{split}
S_{k}^{q}=\lambda_{kq}\theta_{k}B_{q}^{k},
\end{split}
\end{equation}
where the $\theta_{k}$ are the Stevens multiplicative factors for Sm$^{3+}$ listed in Table VI of Ref. \cite{Hutchings}. The constants $\lambda_{20}=\tfrac{1}{2}$ and $\lambda_{40}=\tfrac{1}{8}$, used below to relate our CEF parameters to those calculated by Mauws \emph{et al.}, are listed in Ref. \cite{spectre}.

The double differential neutron cross section \cite{PhysRevB.91.224430}
\begin{eqnarray}\label{xsec}
\frac{d^2\sigma}{d\Omega dE} \propto  \frac{k}{k'} F^2(|\mathbf{Q}|)e^{-2W(\mathbf{Q})} \sum_n p_n \sum_m |\langle m | \hat{M}_\perp(\mathbf{Q}) | n \rangle | ^2 \nonumber \\ 
\times\ \delta(E_m-E_n-E) \phantom{\sum}
\end{eqnarray}
determines the information that can be accessed via neutron scattering. In Eq. (\ref{xsec}), $k$ and $k'$ are respectively the incident and scattered wave-vectors, $F^2(|\mathbf Q|)$ is the magnetic form-factor squared, and $\ket{m},\ket{n}$ are the final and initial CEF states. The population of the $n^{th}$ CEF level is $p_n$ and $e^{-2W(\mathbf{Q})}$ is the Debye-Waller factor. In this work, we consider $W(\mathbf{Q})=0$. This is common assumption in the literature \cite{PhysRevB.91.224430,0953-8984-24-25-256003}, despite being rigorously correct only at $\mathbf{Q}=0$. The component of the magnetisation perpendicular to the scattering vector $\mathbf{Q}$ is denoted by $\hat{M}_\perp(\mathbf{Q})$. When intermultiplet transitions are considered, $\langle \hat{M}_\perp(\mathbf{Q}) \rangle=-\tfrac{2}{3}\langle \hat{L}_\alpha+2\hat{S}_\alpha \rangle$ \cite{OSBORN19911}, where $\hat{L}$ and $\hat{S}$ are the orbital and spin angular momentum operators and $\alpha$ is any of the orthogonal $x,y,z$-axis. The factor $\tfrac{2}{3}$ appears due to the powder averaging of Eq. (\ref{xsec}) \cite{abragam}.

\begin{figure}
\includegraphics[trim={0cm 1.1cm 0cm 0cm},clip,width=8.5cm]{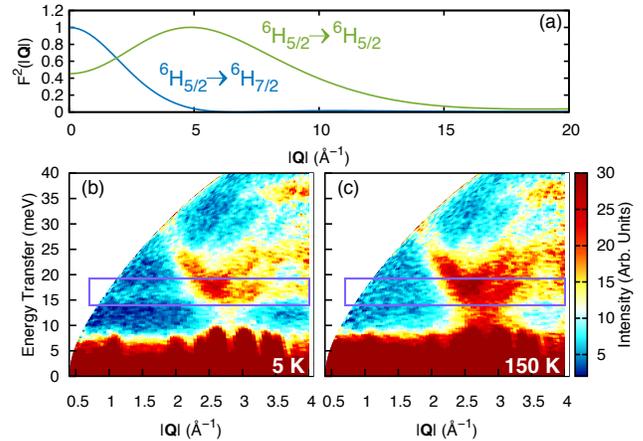}
\caption{(a) Calculated $F^2(|\mathbf Q|)$, in the dipole approximation, for transitions between the same ${^6}H_{5/2}$ multiplet (green) and from the ground-state to the first excited multiplet ${^6}H_{7/2}$ (blue) of the Sm$^{3+}$ ion. The maxima of both curves were normalised to be identically equal to one. (b)-(c) Contour color plots of $\mathrm{^{154}Sm_2Ti_2O_7}$ data measured at MERLIN with a neutron initial energy $E_i=50\ \mathrm{meV}$ at (b) $5\ \textup{K}$ and (c) $150\ \textup{K}$. The purple rectangles highlight the position of one possible CEF excitation at $\sim 16\ \textup{meV}$.}\label{fig:Ei50meV_Sm}
\end{figure}

\begin{figure}
\includegraphics[width=6.0cm,angle=270]{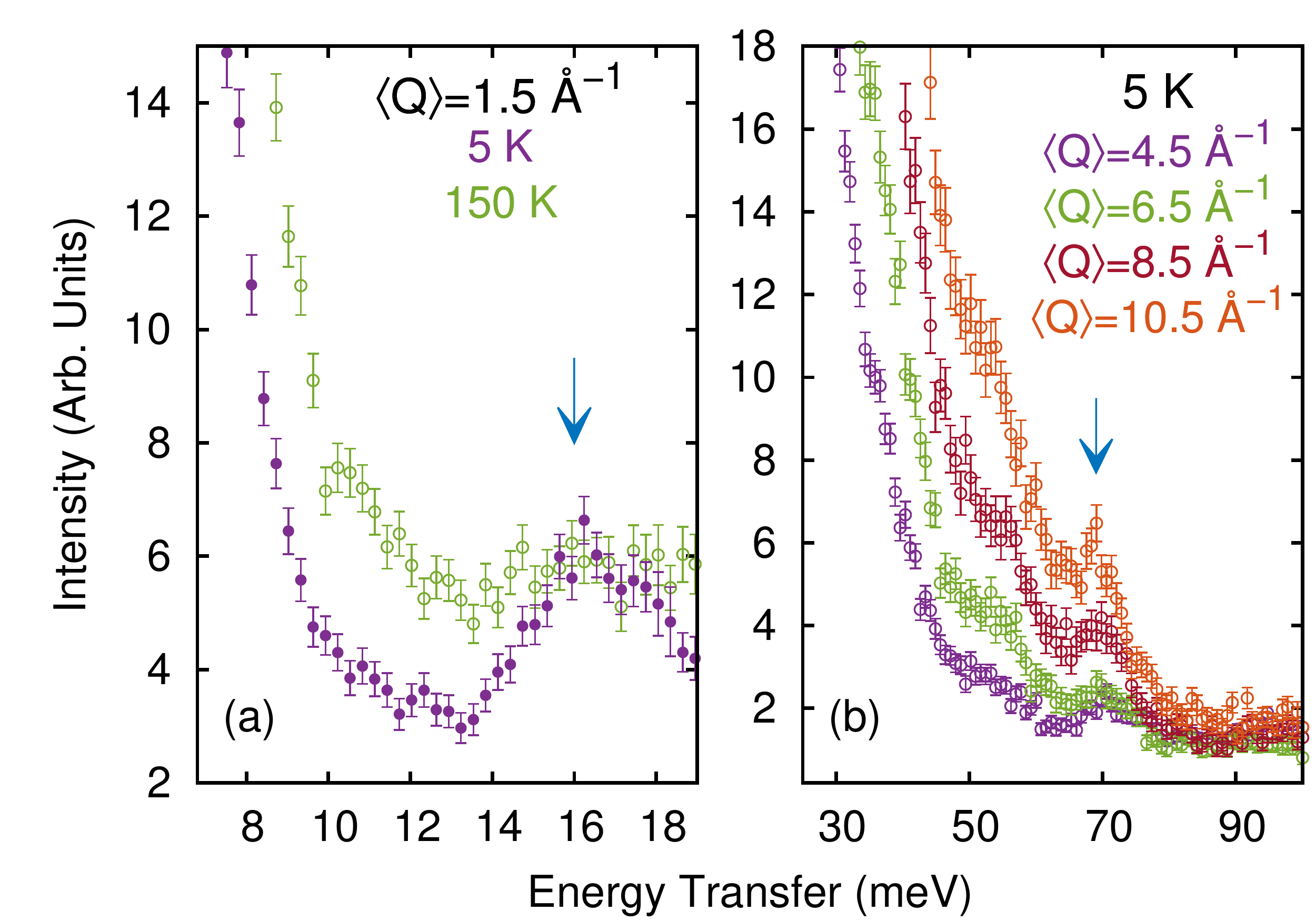}
\caption{Momentum transfer cuts, performed on data shown in Fig. \ref{fig:Ei50meV_Sm}\hyperref[fig:cuts_ground]{(a)} and \hyperref[fig:cuts_ground]{(b)}, integrated over the interval $\langle\mathbf Q\rangle\pm 0.5\ \textup{\AA}^{-1}$. (a) Temperature dependence of the mode observed at $\sim16\ \textup{meV}$. (b) $|\mathbf Q|$-dependence of the flat excitation measured at $70\ \textup{meV}$. Note that the intensity of the excitation in (b) does not follow the form-factor calculated for transitions ${^6}H_{5/2}\leftrightarrow{^6}H_{5/2}$ [see Fig. \ref{fig:Ei50meV_Sm}\hyperref[fig:Ei50meV_Sm]{(a)}].}
\label{fig:cuts_ground}
\end{figure}

Fig. \ref{fig:Ei50meV_Sm}\hyperref[fig:Ei50meV_Sm]{(a)} shows the $F^2(|\mathbf{Q}|)$ of the Sm$^{3+}$ ion \cite{OSBORN19911}. For the transitions within the ground-state multiplet, labeled as ${^6}H_{5/2}\leftrightarrow{^6}H_{5/2}$, the curve is peaked at around $5\ \textup{\AA}^{-1}$, instead of having the (most common) maximum at zero and falling with increasing momentum transfer. In Fig. \ref{fig:Ei50meV_Sm}, panels \hyperref[fig:Ei50meV_Sm]{(b)} and \hyperref[fig:Ei50meV_Sm]{(c)}, the data measured at MERLIN with neutrons of incident energy $E_i=50\ \textup{meV}$ are presented. It is clear that a reliable separation of CEF excitations and flat phonon branches in the experimentally accessed $|\mathbf Q|<10\ \textup{\AA}^{-1}$ momentum transfer range is hindered by the particular $F^2(|\mathbf{Q}|)$ behaviour. We consider, therefore, the temperature dependence of our data in order to identify the modes belonging to the ground-state multiplet splitting of the compound. At $5\ \textup{K}$ [panel \hyperref[fig:Ei50meV_Sm]{(b)}], one weak, dispersionless mode can be observed around $16\ \textup{meV}$ in the interval $|\mathbf Q|=[1,2]\ \textup{\AA}^{-1}$. At 150 K [panel \hyperref[fig:Ei50meV_Sm]{(c)}], the mode intensity is surpassed by a general increase in the background, possibly caused by the growing population of pyrochlore optical phonon levels at medium energies. $|\mathbf Q|$-cuts showing the behaviour of this excitation with increasing temperature are shown in Fig. \ref{fig:cuts_ground}\hyperref[fig:cuts_ground]{(a)}. 

Singh \emph{et al.} \cite{PhysRevB.77.054408} reports, at energy transfers $\Delta E<60\ \mathrm{meV}$, six expected Raman-active phonon modes and other four supposed CEF modes lying at energies around 11, 16, 20 and $33\ \textup{meV}$. While we identify an excitation $\sim16\ \mathrm{meV}$, our data measured with $E_i=19\ \mathrm{meV}$ neutrons (not shown) display no evidence of a CEF level at $11\ \textup{meV}$. Similarly, the high density of phonon states around 20 and $33\ \textup{meV}$ does not allow us to directly identify, in the measured $S(|\mathbf Q |,\omega)$, those higher-energy, supposed single-ion excitations. Below, we are going to show that a mode at $\sim 30\ \textup{meV}$ is predicted by our CEF analysis. The levels located at 11 and $20\ \textup{meV}$ were suggested, in an earlier Raman and IR spectroscopy work performed on $\mathrm{Sm_2Ti_2O_7}$ \cite{vandeborre}, to correspond to two IR-active phonon levels, which become also Raman-active due to a local symmetry lowering associated with crystal defects. 

The inelastic neutron scattering data of Mauws \emph{et al.} \cite{PhysRevB.98.100401} indicate a CEF excitation taking place at $\sim16\ \textup{meV}$ and one additional at $70\ \textup{meV}$. Indeed, we measure a flat excitation at $70\ \textup{meV}$ using neutrons with $E_i=150\ \mathrm{meV}$. However, $|\mathbf Q|$-cuts performed for several momentum transfer intervals, displayed in Fig. \ref{fig:cuts_ground}\hyperref[fig:cuts_ground]{(b)}, reveal that this mode intensity increases with increasing $\langle\mathbf Q\rangle$ up to at least $10.5\ \textup{\AA}^{-1}$. This behaviour clearly contradicts the expected considering the form-factor for transitions ${^6}H_{5/2}\leftrightarrow{^6}H_{5/2}$, as shown in Fig. \ref{fig:Ei50meV_Sm}\hyperref[fig:Ei50meV_Sm]{(a)}. We also note that this position in energy coincides with those of several $\Gamma$-point phonons in pyrochlores \cite{PhysRevB.93.214308}. 

As the observation of one (or two) CEF levels is not usually enough to perform a reliable calculation of the single-ion Hamiltonian, we continue our search for crystal-field excitations at higher energies. Following a trend common also in other light rare-earth compounds, the ground-state ${^6}H_{5/2}$ multiplet of Sm is relatively close in energy to the first excited ${^6}H_{7/2}$ state, and its splitting into four doublets is expected to be observed at energies below 200 meV \cite{Carnall}. 


\begin{figure}
\includegraphics[trim={2.5cm 0cm 2.5cm 0cm},clip,width=4.2cm,angle=270]{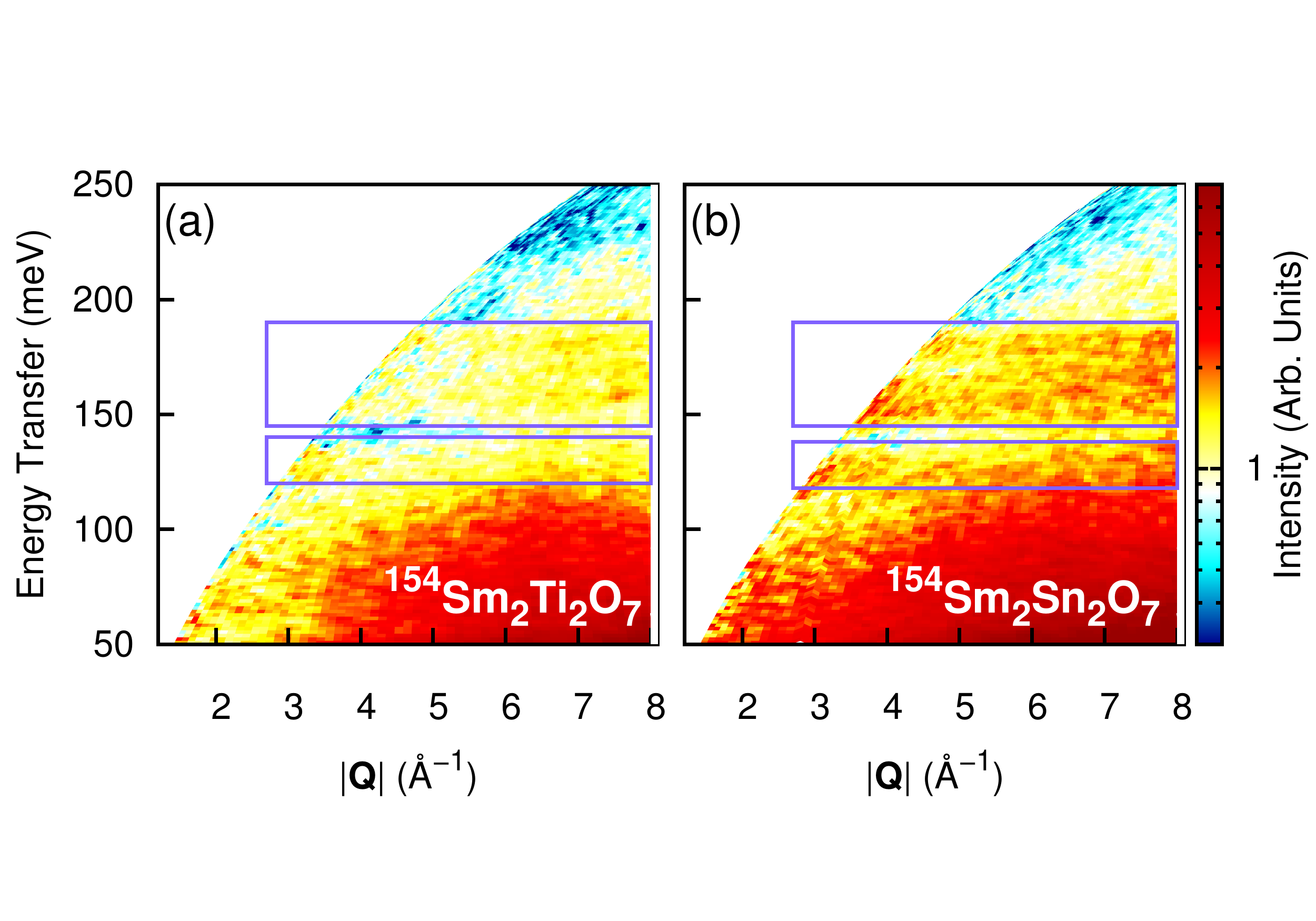}
\caption{Possible CEF excitations, emphasised by blue rectangles, resulting from the splitting of the first excited multiplet on (a) titanate and (b) stannate samples. Data were collected at 5 K using neutrons of incident energy $E_i=300\ \mathrm{meV}$.}
\label{fig:Ei300meV_Sm}
\end{figure}

\begin{figure}
\centering
\includegraphics[trim={0cm 5.25cm 0cm 7.1cm},clip,width=9cm,angle=270]{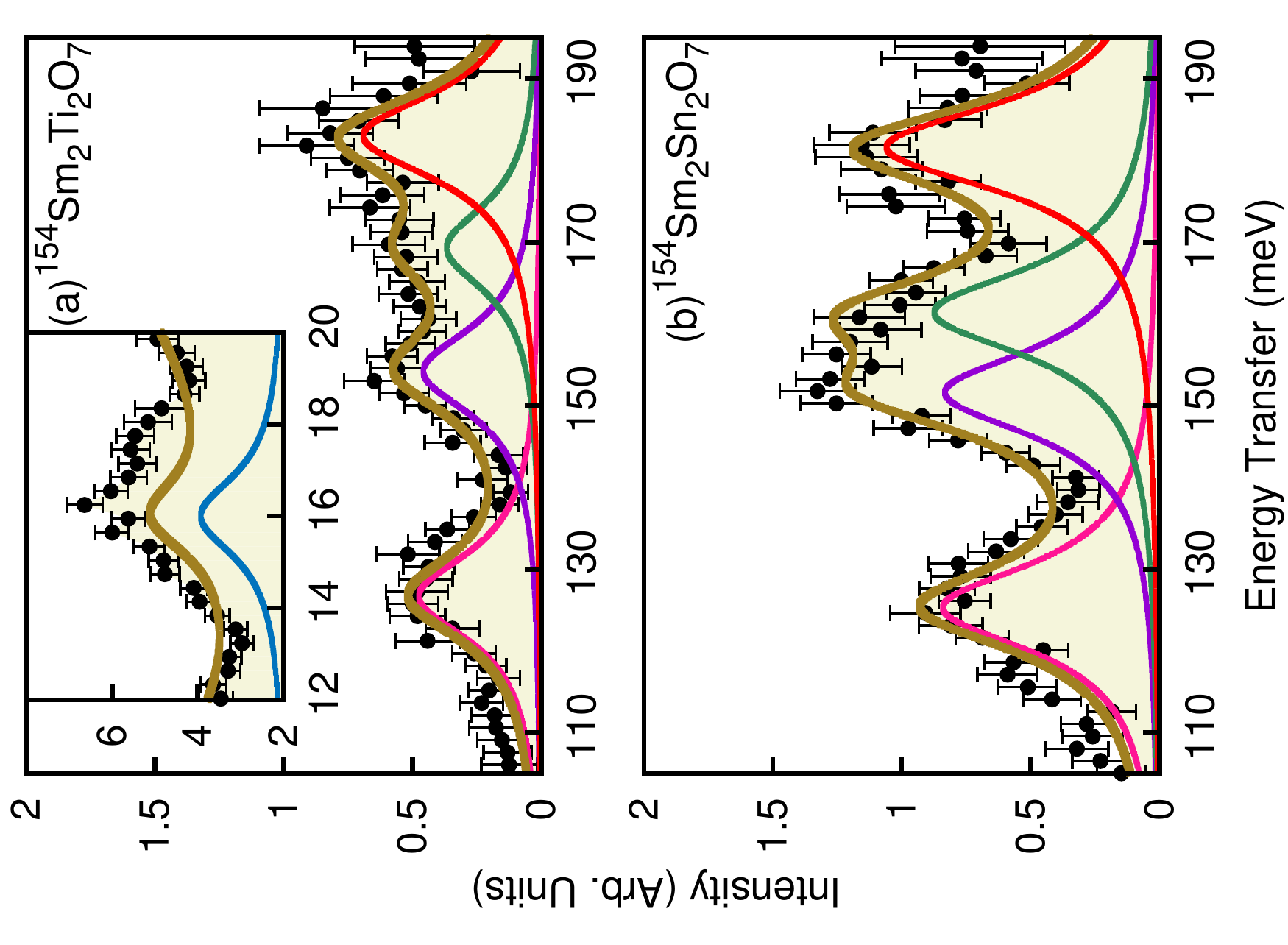}
\caption{Data (black circles, with error bars) with superimposed calculated intensities (continuous lines) obtained in our CEF analysis (see Table \ref{tab:tab8}) for (a) titanate and (b) stannate. Note that the intensity of the level at $16\ \textup{meV}$ was not used in the fitting, nevertheless we show for completeness its calculated intensity in the inset of (a).}
\label{fig:cuts}
\end{figure}

In Fig. \ref{fig:Ei300meV_Sm}\hyperref[fig:Ei300meV_Sm]{(a)} and \hyperref[fig:Ei300meV_Sm]{(b)}, the spectra measured at 5 K with $E_i=300\ \mathrm{meV}$ neutrons are shown for $\mathrm{^{154}Sm_2Ti_2O_7}$ and $\mathrm{^{154}Sm_2Sn_2O_7}$, respectively. Despite the weak signal, some low-$|\mathbf Q|$ flat modes are present in both samples. We perform $|\mathbf Q|$-cuts in the data of Fig. \ref{fig:Ei300meV_Sm}, integrating the intensity over the full interval where the form-factor for transitions ${^6}H_{5/2}\leftrightarrow{^6}H_{7/2}$ is appreciably bigger than zero, i.e. where $|\mathbf Q|<5\ \textup{\AA}^{-1}$. Those cuts are fit with Lorentzian distributions centred on the peak position. In order to define the relative intensities of the modes, the FWHM of the excitations were assumed to be identical, which is a reasonable approximation for energies between 120 and 190 meV. For the background estimation, we performed higher momentum transfer cuts with same width in $|\mathbf Q|$. The intensities of both were scaled $\sim100\ \mathrm{meV}$, and, after subtraction, the data shown in Fig. \ref{fig:cuts} (filled black circles) are obtained. The parameters resulting from the Lorentzian-fittings are summarised in Table \ref{tab:tab7}.


\begin{table}
\tabcolsep=2.5pt
\renewcommand{\arraystretch}{1}
\begin{tabular}{c c c c c c c c}
\toprule[1pt]
\multicolumn{8}{c }{$\mathrm{Sm_2Ti_2O_7}$}\\
\cmidrule[0.3pt]{1-8}
\multicolumn{2}{c }{}&\multicolumn{4}{c }{Observed}&\multicolumn{2}{c }{Calculated}\\
\cmidrule[0.3pt](r{.75em}){3-6}\cmidrule[0.3pt]{7-8}
\multicolumn{2}{c }{Transition}	&	Energy	&	$\sigma_E$	&	Relative	&	$\sigma_I$	&	Energy	&	Relative 	\\
\multicolumn{2}{c }{}	&	(meV)	&	(meV)	&	Intensity	&	&	(meV)	&	Intensity	\\
\cmidrule[0.3pt](r{.75em}){1-6}\cmidrule[0.3pt]{7-8}
\multicolumn{2}{c }{$\ket{0}\leftrightarrow\ket{0}$}	&	0	&	0	&	-	&	-	&	0	&	0.05	\\
\multicolumn{2}{c }{$\ket{0}\leftrightarrow\ket{1}$}	&	16.1*	&	0.2	&	-	&	-	&	16.0	&	0.15	\\
\multicolumn{2}{c }{$\ket{0}\leftrightarrow\ket{2}$}	&	- 	&	-	&	-	&	-	&	29.7	&	0.02	\\
\multicolumn{2}{c }{$\ket{0}\leftrightarrow\ket{3}$}	&	126.7	&	0.5	&	1	&	- 	&	127.2	&	1	\\
\multicolumn{2}{c }{$\ket{0}\leftrightarrow\ket{4}$}	&	154.2	&	0.5	&	1.0	&	0.1	&	154.0	&	0.74	\\
\multicolumn{2}{c }{$\ket{0}\leftrightarrow\ket{5}$}	&	169.3	&	0.9	&	0.8	&	0.1	&	169.6	&	0.92	\\
\multicolumn{2}{c }{$\ket{0}\leftrightarrow\ket{6}$}	&	183.0 	&	0.6	&	1.5	&	0.1	&	183.4	&	1.48	\\
\bottomrule[0.5pt]
\multicolumn{8}{c }{$\mathrm{Sm_2Sn_2O_7}$}\\
\cmidrule[0.3pt]{1-8}
\multicolumn{2}{c }{}&\multicolumn{4}{c }{Observed}&\multicolumn{2}{c }{Calculated}\\
\cmidrule[0.3pt](r{.75em}){3-6}\cmidrule[0.3pt]{7-8}
\multicolumn{2}{c }{Transition}	&	Energy	&	$\sigma_E$	&	Relative	&	$\sigma_I$	&	Energy	&	Relative 	\\
\multicolumn{2}{c }{}	&	(meV)	&	(meV)	&	Intensity	&	&	(meV)	&	Intensity	\\
\cmidrule[0.3pt](r{.75em}){1-6}\cmidrule[0.3pt]{7-8}
\multicolumn{2}{c }{$\ket{0}\leftrightarrow\ket{0}$}	&	0	&	0	&	-	&	-	&	0	&	0.05	\\
\multicolumn{2}{c }{$\ket{0}\leftrightarrow\ket{1}$}	&	-	&	-	&	-	&	-	&	13.0	&	0.18	\\
\multicolumn{2}{c }{$\ket{0}\leftrightarrow\ket{2}$}	&	- 	&	-	&	-	&	-	&	28.2	&	0.01	\\
\multicolumn{2}{c }{$\ket{0}\leftrightarrow\ket{3}$}	&	125.2	&	0.4	&	1	&	- 	&	125.2	&	1	\\
\multicolumn{2}{c }{$\ket{0}\leftrightarrow\ket{4}$}	&	151.6	&	0.4	&	1.1	&	0.1	&	151.6	&	1.02	\\
\multicolumn{2}{c }{$\ket{0}\leftrightarrow\ket{5}$}	&	161.4	&	0.5	&	0.9	&	0.1	&	161.4	&	1.04	\\
\multicolumn{2}{c }{$\ket{0}\leftrightarrow\ket{6}$}	&	181.6 	&	0.5	&	1.3	&	0.1	&	181.6	&	1.23	\\
\bottomrule[0.5pt]
\end{tabular}
\caption{Overview of the observed and calculated eigenstates and intensities of the CEF levels of $\mathrm{^{154}Sm_2Ti_2O_7}$ and $\mathrm{^{154}Sm_2Sn_2O_7}$. The position of the levels and their intensities (except for the level marked with *, which did not have its intensity used in the calculation) were obtained by fitting data shown in Fig. \ref{fig:cuts_ground}\hyperref[fig:cuts_ground]{(a)} and Fig. \ref{fig:cuts}. The $\sigma$ values are the standard deviation of the Lorentzian fittings. The $\ket{n}$ represents both degenerated wave-functions of the $n^{th}$ excited doublet. The intensities of the second multiplet splitting were all calculated relative to the transition $\ket{0}\leftrightarrow \ket{3}$.}\label{tab:tab7}
\end{table}

To confirm that those modes really correspond to CEF levels, the Hamiltonian of Eq. (\ref{eq:eq1}) is fit to the eigenenergies and intensity ratios obtained experimentally. The inclusion of intermultiplet transitions in the model precludes the use of the Stevens' operator formalism in the analysis of the CEF excitations in both compounds. The only approach left to treat the problem is then to use the Wybourne tensor operator formalism and diagonalise the resulting $H_{CEF}$ Hamiltonian. To perform this task we make use of the software {{\normalsize S}{\footnotesize PECTRE}} \cite{spectre}. The complete $4f^5$ basis of states of Sm$^{3+}$ is reduced to comprise only the $12\ {J}$ values belonging to the two lowest multiplets in energy ${^6}H$ and ${^6}F$ of the ion \cite{Carnall}. Intermediate coupling, which is the mixing of levels with the same $J$ but different $L$ and $S$ quantum numbers \cite{wybourne}, was also taken into account. The starting parameters were taken from Malkin \emph{et al.} \cite{Malkin}.

Fig. \ref{fig:cuts}\hyperref[fig:cuts]{(a)} and \hyperref[fig:cuts]{(b)} display, for titanate and stannate, respectively, the calculated position and intensity of the best-fit single-ion CEF model (continuous golden line). For immediate comparison with experiment, those results are also shown in Table \ref{tab:tab7}. The energy of the mode at $16\ \mathrm{meV}$ was considered in the titanate CEF model. However, our attempt to include the mode at $70\ \textup{meV}$ in a fitting comprising the full dataset obtained for this compound was unsuccessful. We thus rule out the possibility that this excitation corresponds to a CEF level.

Our calculations predict that the unmeasured second excited level of the ground-state multiplet of titanate and stannate should be present at $\sim30\ \textup{meV}$. This is fully consistent with the Raman spectroscopy results of Ref. \cite{PhysRevB.77.054408}. The estimated intensity of the transitions $\ket{0}\leftrightarrow\ket{2}$ is the smallest in Table \ref{tab:tab7}. That suggests another cause, besides the massive phonon density of states at intermediate energies, for the seemingly absence of this mode in the experimental $S(|\mathbf Q |,\omega)$. The eigenenergies of the $H_\mathrm{CEF}$ of stannate and titanate are very similar, with the strongest differences lying on the modes relative intensities. In the inelastic neutron scattering work performed in the zirconate \cite{xu_jianhui}, CEF levels were measured $\sim130$, 157, 168 and $183\ \mathrm{meV}$, in close agreement with the samples analysed here.   


The calculated crystal field parameters are shown in Table \ref{tab:tab8}. The values given in Ref. \cite{PhysRevB.98.100401} were converted to the Wybourne tensor formalism using Eq. (\ref{eq2}), and are displayed in the same Table for comparison. As a direct consequence of the use of the Stevens' formalism, the non-diagonal elements in the Hamiltonian matrix obtained in Mauws \emph{et al.} \cite{PhysRevB.98.100401} (with exception of $B_{3}^{4}$) are identically equal to zero. 

\begin{table}
\centering
\tabcolsep=4pt
\renewcommand{\arraystretch}{1.2}
\begin{tabular}{c c c c c c c}
\toprule[1pt]
	&	$B_{0}^{2}$	&	$B_{0}^{4}$	&	$B_{3}^{4}$	&	$B_{0}^{6}$	&	$B_{3}^{6}$	&	$B_{3}^{6}$	\\
\cmidrule[0.3pt]{2-7}
$\mathrm{Sm_2Ti_2O_7}$	&	73.7	&	369.5	&	102.5	&	167.1	&	-123.0	&	141.8	\\
$\mathrm{Sm_2Sn_2O_7}$	&	83.1	&	319.7	&	111.8	&	133.4	&	-110.8	&	155.2	\\
\cmidrule[0.3pt]{1-7}
$\mathrm{Sm_2Ti_2O_7}$ \cite{Malkin}	&	28.5	&	370.0	&	97.3	&	87.0	&	-78.0	&	124.0	\\
$\mathrm{Sm_2Ti_2O_7}$ \cite{PhysRevB.98.100401}	&	164.6	&	393.4	&	0	&	0	&	0	&	0	\\
\bottomrule[0.5pt]
\end{tabular}
\caption{Crystal field parameters, in units of meV, obtained in this work. The two last lines in the Table show, for comparison, the set of crystal field parameters published in the Refs. \cite{PhysRevB.98.100401,Malkin}. Note that the parameters of Ref. \cite{PhysRevB.98.100401} are estimated using the Stevens' operator formalism, with the immediate consequence that all $B_{q}^{6}$ parameters are equal to zero.}\label{tab:tab8}
\end{table}

\begin{table*}
\centering
\tabcolsep=7pt
\renewcommand{\arraystretch}{1.5}
\begin{tabular}{c c}
\toprule[1pt]
$\mathrm{Sm_2Ti_2O_7}$	&$\ket{\pm0}=\pm0.303\ket{{^6}H_{5/2},\mp\frac{3}{2}} + 0.934\ket{{^6}H_{5/2},\pm\frac{3}{2}}+ 0.016\ket{{^6}H_{7/2},\mp\frac{3}{2}} \mp 0.073\ket{{^6}H_{7/2},\pm\frac{3}{2}}$ \\
						&$\mp0.030\ket{{^6}H_{9/2},\mp\frac{3}{2}}-0.084\ket{{^6}H_{9/2},\pm\frac{3}{2}}+0.093\ket{{^6}H_{9/2},\mp\frac{9}{2}}\mp0.017\ket{{^6}H_{9/2},\pm\frac{9}{2}}$		\\
\cmidrule[0.2pt](l{.75em}r{.75em}){1-2}
$\mathrm{Sm_2Sn_2O_7}$	&$\ket{\pm0}=\pm0.084\ket{{^6}H_{5/2},\mp\frac{3}{2}} + 0.981\ket{{^6}H_{5/2},\pm\frac{3}{2}}+ 0.005\ket{{^6}H_{7/2},\mp\frac{3}{2}} \mp 0.058\ket{{^6}H_{7/2},\pm\frac{3}{2}}$ \\
						&$\mp0.012\ket{{^6}H_{9/2},\mp\frac{3}{2}}-0.076\ket{{^6}H_{9/2},\pm\frac{3}{2}}+0.093\ket{{^6}H_{9/2},\mp\frac{9}{2}}\mp0.039\ket{{^6}H_{9/2},\pm\frac{9}{2}}$		\\
\cmidrule[0.2pt](l{.75em}r{.75em}){1-2}
$\mathrm{Sm_2Ti_2O_7}$ \cite{Malkin}	&	$\ket{\pm0}=\pm 0.154\ket{{^6}H_{5/2},\mp\frac{3}{2}} + 0.970\ket{{^6}H_{5/2},\pm\frac{3}{2}}+ 0.028\ket{{^6}H_{7/2},\mp\frac{3}{2}}\mp 0.125\ket{{^6}H_{7/2},\pm\frac{3}{2}}$ \\
						&$\mp0.016\ket{{^6}H_{9/2},\mp\frac{3}{2}}-0.045\ket{{^6}H_{9/2},\pm\frac{3}{2}}+0.078\ket{{^6}H_{9/2},\mp\frac{9}{2}}\mp0.023\ket{{^6}H_{9/2},\pm\frac{9}{2}}$		\\
\cmidrule[0.2pt](l{.75em}r{.75em}){1-2}
$\mathrm{Sm_2Ti_2O_7}$ \cite{PhysRevB.98.100401}	&	$\ket{\pm0}=\ket{{^6}H_{5/2},\mp\frac{3}{2}}$\\
\bottomrule[0.5pt]
\end{tabular}
\caption{Ground-state wave-functions of titanate and stannate. We use the spectroscopic notation $^{2S+1}L_{J}$ followed by the magnetic quantum number $m_J$ to assign each basis functions $|L,S,J,m_J\rangle$. The bottom lines in the Table are dedicated to the values found in the analysis of Refs. \cite{PhysRevB.98.100401,Malkin}. The wave-function of Malkin \emph{et al.} \cite{Malkin} was not reported in the original publication, so we recalculate it here based on the parameters obtained and reproduced in Table \ref{tab:tab8}. }\label{tab:tab9}
\end{table*}

The calculated doublet ground-state wave-functions $\ket{\pm0}$ are shown in Table \ref{tab:tab9}, which contains also the $\ket{\pm0}$ states of Refs. \cite{PhysRevB.98.100401,Malkin}. In general, all of them are comprised mostly by the $\ket{{^6}H_{5/2},\mp\frac{3}{2}}$ state, with contributions from the higher $J=7/2$ and $J=9/2$ multiplets. The ground-state, dipolar magnetic moment of the $\mathrm{Sm^{3+}}$ is determined using 
\begin{equation}
\label{eq:eq5}
|\langle \hat\mu^{0}_\alpha \rangle|=\vert \langle \pm 0 \vert-(\hat L_\alpha+2\hat S_\alpha )\mu_\mathrm{B} \vert \pm 0\rangle \vert. 
\end{equation}
The only matrix elements different of zero in Eq. (\ref{eq:eq5}) are those for which $\alpha=z$, presented in Table \ref{tab:tab10}. The ground-state anisotropy of the samarium pyrochlores is thus Ising-like, or, conversely, the crystal field constrains the magnetic moment to point along one of the crystallographic $\langle 111 \rangle$ axis, or local $z$-axis.  

Even though the admixture of ${^6}H_{7/2}$ and ${^6}H_{9/2}$ terms in the ground-state doublet is minimal, its effect over the ground-state magnetic moment is surprisingly important: $\langle \hat\mu^{0}_z \rangle$ changes from the maximum $0.43\mu_\mathrm{B}$ for a pure $\ket{{^6}H_{5/2},\mp\frac{3}{2}}$ doublet, the result of Ref. \cite{PhysRevB.98.100401}, to the $0.16\mu_\mathrm{B}$ expected in our work for the titanate or $0.27\mu_\mathrm{B}$ for the stannate. In anticipation of our low-temperature neutron diffraction results, and focusing specifically on the titanate, the reduction of the magnetic moment from $0.43\mu_\mathrm{B}$ to $0.16\mu_\mathrm{B}$ transforms the rather small, but measurable $\langle \hat\mu^{0}_z \rangle$ of $\mathrm{^{154}Sm_2Ti_2O_7}$ into a virtually undetectable quantity.

\begin{table}[h]
\tabcolsep=20pt
\renewcommand{\arraystretch}{1.2}
\begin{tabular}{c c}
\toprule[1pt]
	&	$|\langle\hat \mu^{\pm0}_\alpha\rangle|=|\langle\hat \mu^{\pm0}_z\rangle|$ ($\mu_\mathrm{B}$) \\
\cmidrule[0.3pt](l{.75em}r{.75em}){1-2}
$\mathrm{Sm_2Ti_2O_7}$	&	$0.16$	\\
$\mathrm{Sm_2Sn_2O_7}$	&	$0.27$	\\
\cmidrule[0.2pt](l{.75em}r{.75em}){1-2}
$\mathrm{Sm_2Ti_2O_7}$ \cite{Malkin}	&	$0.11$\\
$\mathrm{Sm_2Ti_2O_7}$ \cite{PhysRevB.98.100401}	&	$0.43$\\
\bottomrule[0.5pt]
\end{tabular}
\caption{Ground-state CEF magnetic moments calculated in this work and the ones found by Mauws \emph{et al.} \cite{PhysRevB.98.100401} and Malkin \emph{et al.} \cite{Malkin}.} \label{tab:tab10}
\end{table}

\subsection{Magnetic susceptibility}

\begin{figure}
\centering
\includegraphics[trim={3.3cm 0.5cm 3.5cm 0.7cm},clip,width=3.9cm,angle=270]{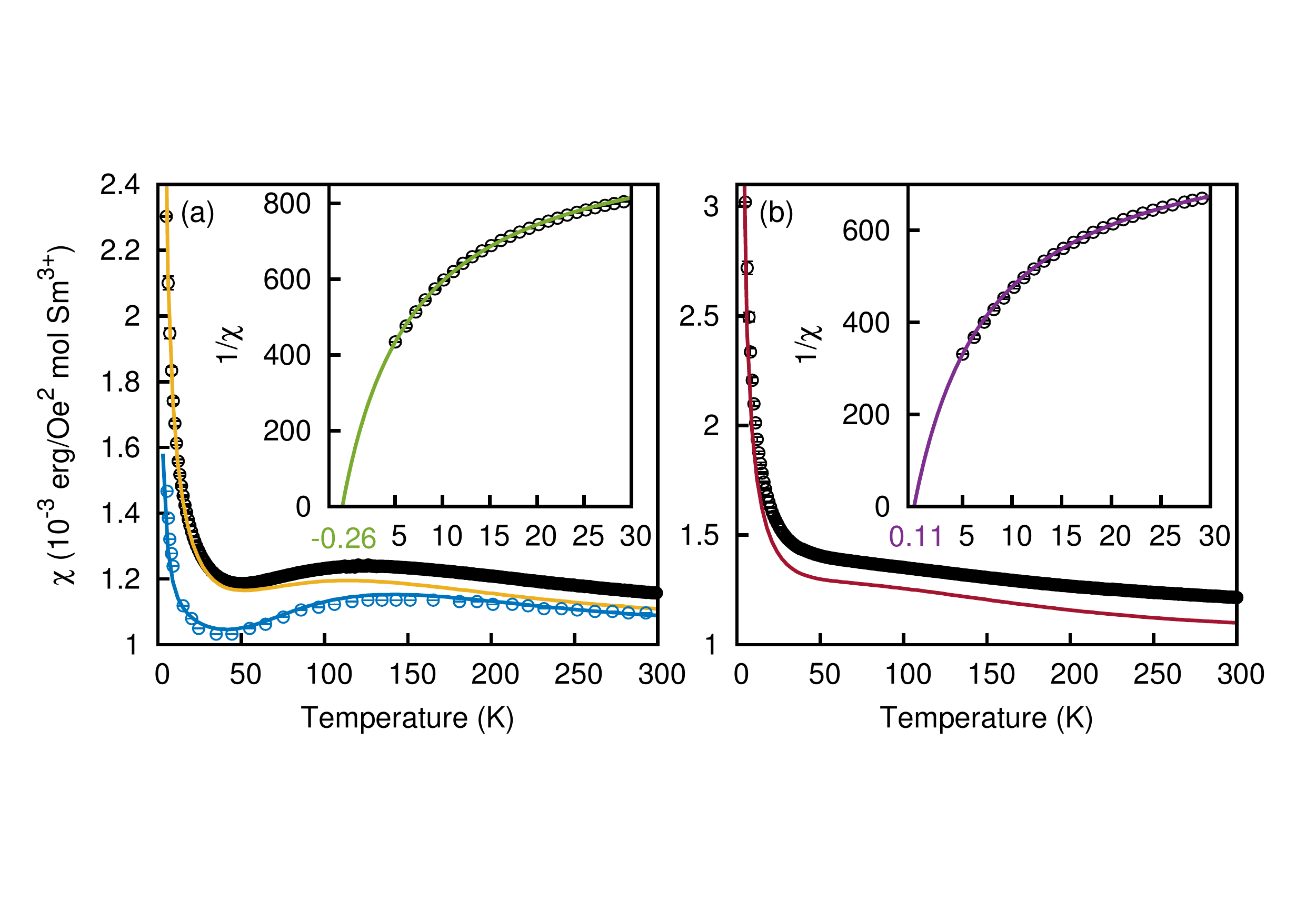}
\caption{Powder-averaged, static magnetic susceptibility $\chi_\mathrm{dc}$ measured at 1000 Oe applied magnetic field for (a)$\mathrm{^{154}Sm_2Ti_2O_7}$ and (b)$\mathrm{^{154}Sm_2Sn_2O_7}$. The continuos lines show the $\chi_{\mathsmaller{CEF}}$ calculated using Eq. (\ref{eq3}). The blue circles in (a) reproduce data of Singh \emph{et al.} \cite{PhysRevB.77.054408} and the blue line corresponds to the fitting of Malkin \emph{et al.} \cite{Malkin}. The insets in (a) and (b) show the Curie-Weiss fittings of the reciprocal susceptibility, performed using Eq. (\ref{eq:eq6}).}
\label{fig:suscept_154Sm}
\end{figure}

The susceptibility $\chi_\mathrm{dc}$ data are shown in Fig. \ref{fig:suscept_154Sm}\hyperref[fig:suscept_154Sm]{(a)} and \hyperref[fig:suscept_154Sm]{(b)}, respectively for titanate and stannate. The $\chi_\mathrm{dc}$ of $\mathrm{^{154}Sm_2Ti_2O_7}$ is easily recognisable, for it shows a dip just below 50 K followed by a maximum at $\sim120\ \mathrm{K}$, after which it decreases monotonically up to room temperature. The susceptibility of $\mathrm{^{154}Sm_2Sn_2O_7}$, on the other hand, displays no maximum up to 300 K. 

To perform the calculation of the single-ion contribution to the magnetic susceptibility, we make use of the van Vleck equation \cite{vanvleck}
\begin{eqnarray}
\label{eq3}
 \chi_{{\mathsmaller{CEF}},\alpha}=\frac{2N\beta}{Z} \left[\sum_{n}{\frac{\vert\langle {+n} \vert \hat\mu_\alpha \vert {+n} \rangle\vert^2+\vert\langle +n \vert \hat\mu_\alpha\vert {-n} \rangle\vert^2}{e^{\beta{E_n}}}} \right. &\nonumber \\
\left. +\ \ \frac{2}{\beta}\sum_{m> n }\frac{\vert\langle {+m} \vert \hat\mu_\alpha \vert {+n} \rangle\vert^2+\vert\langle {+m} \vert \hat\mu_\alpha \vert {-n} \rangle\vert^2}{E_n-E_m}\right. &\nonumber \\
\left. \phantom{\sum_{m> n\ge0 }}\times (e^{-\beta{E_m}}-e^{-\beta{E_n}})\right], \qquad&
\end{eqnarray}
where $N$ is the Avogadro's number, $Z=\sum_{n} {e^{-\beta E_n}}$ is the partition function and $\beta=1/k_BT$. $E_{n,m}$ are the eigenvalues of the CEF Hamiltonian, presented in Table \ref{tab:tab7}, and $k_B$ is the Boltzmann constant. The magnetic moment operator $\hat\mu_\alpha$ was defined in Eq. (\ref{eq:eq5}). All the terms for which $m>n$ [second summation on the right hand side of Eq. (\ref{eq3})] correspond to the so-called \emph{van Vleck} susceptibility. Note that we neglect the sample own diamagnetic response. The powder $\chi_{\mathsmaller{CEF}}$, calculated averaging Eq. (\ref{eq3}) over $x,y,z$ directions, is shown along with the experimental results in Fig. \ref{fig:suscept_154Sm}\hyperref[fig:suscept_154Sm]{(a)} and \hyperref[fig:suscept_154Sm]{(b)}.

Our main objective with the single-ion susceptibility calculation is to ratify our CEF analysis, particularly for the stannate, since the ground-state splitting of the $\mathrm{Sm^{3+}}$ single-ion multiplet was not observed. The absolute values of $\chi_\mathrm{dc}$ are rather small, of the order of $10^{-3}\ \mathrm{erg/(Oe^2\ \ mol\ \ Sm^{3+}})$. Consequently, our measurements may be extremely sensitive to magnetic impurities and to the perturbation induced by small applied magnetic fields. This suggests one possible cause for the small differences between model and measured values. Still, the $\chi_{CEF}$ calculation reproduces extremely well the peculiar shape of the curves, and confirms that the non-linear behaviour of $\chi_\mathrm{dc}$ at intermediate temperatures is related to single-ion properties of the magnetic ion. 

Also in Fig. \ref{fig:suscept_154Sm}\hyperref[fig:suscept_154Sm]{(a)}, data of Singh \emph{et al.} \cite{PhysRevB.77.054408} and calculation performed by Malkin \emph{et al.} \cite{Malkin}, reproduced by us using the crystal field parameters reported in Ref. \cite{Malkin}, are shown. In the work of Ref. \cite{PhysRevB.77.054408}, the magnetic susceptibility was measured in a single crystal sample and no special orientation of it along the applied magnetic field is defined (at least no one is mentioned). There is some discrepancy between the experimental susceptibilities, which increases as the temperature decreases. This can be a consequence, as we already noted, of the presence of a small amount of magnetic impurities in our sample. Another reason could be that, due to the compounds strong anisotropy, the susceptibility reported in Ref. \cite{PhysRevB.77.054408} does not correspond to a powder averaged $\chi_\mathrm{dc}$ and, unlike the procedure followed by Malkin \emph{et al.}, should not be modelled as such. 

\begin{figure}[h]
\includegraphics[trim={5cm 0.0cm 3.5cm 0cm},clip,width=3.0cm,angle=270]{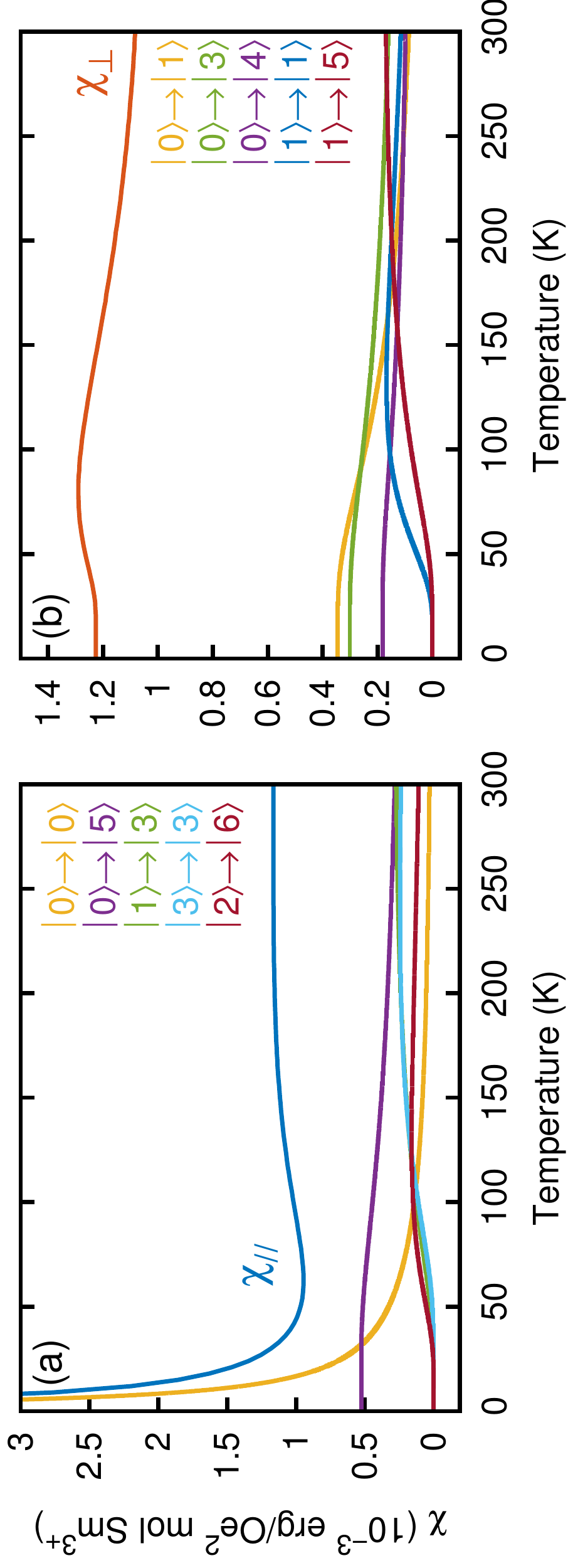}
\caption{Powder susceptibility calculated for $\mathrm{^{154}Sm_2Ti_2O_7}$, separated in (a) parallel $\chi_\parallel$ and (b) perpendicular $\chi_\perp$ components. Each transition $\ket{n}\rightarrow\ket{m}$ represents one term in the summation of Eq. (\ref{eq3}).}
\label{fig:suscept}
\end{figure}

In order to explicitly demonstrate the effect of the increased population of excited CEF levels as the temperature increases, the titanate is taken as example. We split the summations in Eq. (\ref{eq3}) term-by-term, and plot some of them as a function of temperature in Fig. \ref{fig:suscept}. In \hyperref[fig:suscept]{(a)}, the single-ion susceptibility for a field applied along the $z$-axis, or $\chi_\parallel$, is displayed. The component of the magnetic susceptibility perpendicular to the $z$ direction, which we denote by $\chi_\perp$, is given in Fig. \ref{fig:suscept}\hyperref[fig:suscept]{(b)}.

The view of the individual contributions emphasises the striking relative magnitude of the van Vleck susceptibility associated with intermultiplet transitions (note, for example, the $\ket{0}\rightarrow\ket{5}$ along the local $z$-axis). Furthermore, because, around 80 K, $\chi_\parallel$ is minimum where the $\chi_\perp$ is maximum, Fig. \ref{fig:suscept} additionally offers the explanation for the upturn detected in the experimental powder averaged $\chi_\mathrm{dc}$ of the titanate.

Beyond the single-ion suceptibility, we analyse $\chi_\mathrm{dc}$ at low temperatures to obtain informations about the interactions between the magnetic atoms \footnote{In the absence of strong dipolar interactions caused by large magnetic moments, the exchange interaction is the dominant term in the Hamiltonian of pyrochlores \cite{RevModPhys.82.53}.}. The inclusion of exchange interactions into the Hamiltonian of pyrochlores \cite{RevModPhys.82.53} causes a shift $\theta_{CW}$ in the point where the reciprocal susceptibility intercepts the temperature axis \cite{Blundell}. This is modelled by the Curie-Weiss model
\begin{equation}
\chi_{ex}=\frac{N\mu_\mathrm{eff}^2}{3k_B(T-\theta_{CW})}+\chi_{\mathrm{vV}},
\label{eq:eq6}
\end{equation}  
where $\chi_{\mathrm{vV}}$ is the van Vleck susceptibility, $\mu_\mathrm{eff}$ is the effective magnetic moment and $\theta_\mathrm{CW}$ is the Curie-Weiss temperature. 


Eq. (\ref{eq:eq6}) is fit to the susceptibility data shown in Fig. \ref{fig:suscept_154Sm}. At temperatures $T\ll E_{\ket{\pm1}}/k_B\sim150\ \mathrm{K}$, where $E_{\ket{\pm1}}$ is the energy of the first excited CEF doublet in the stannate or titanate, $\chi_{\mathrm{vV}}\sim 10^{-3}\ \mathrm{erg/(Oe^2\ \ mol\ \ Sm^{3+}})$ is expected to be constant. The van Vleck susceptibility was calculated for both compounds using Eq. (\ref{eq3}). Similarly, it could be assumed that, at low temperatures, $\mu_\mathrm{eff}=\langle \hat\mu^{0}_z \rangle$, as given in Table \ref{tab:tab10}. However, since the calculated $\chi_\mathsmaller{CEF}$ does not reproduce optimally the measured susceptibility, $\mu_\mathrm{eff}$ and $\chi_{\mathrm{vV}}$, together with the $\theta_\mathrm{CW}$, are kept as fitting parameters. 


\begin{table}[h]
\tabcolsep=3pt
\renewcommand{\arraystretch}{1.2}
\begin{tabular}{c c c c c c}
\toprule[1pt]
&\multicolumn{2}{c }{$\mathrm{^{154}Sm_2Ti_2O_7}$}&&\multicolumn{2}{c }{$\mathrm{^{154}Sm_2Sn_2O_7}$}\\
\cmidrule[0.3pt]{2-6}
Fitting inteval (K)&$[5,30]$ 	&$[5,20]$	&	&$[5,30]$ &$[5,20]$ \\
\cmidrule[0.3pt]{1-6}
$\mu_\mathrm{eff}\ (\mu_\mathrm{B})$ &0.23(2)&0.24(3)&	&0.27(3)	&0.29(4)\\
$\theta_\mathrm{CW}$ (K) &-0.26(4)	&-0.43(8)&	&0.11(6)	&-0.44(7)\\
$\chi_{\mathrm{vV}}\ (\times10^{-4})$ &10.10(2)&9.98(6)&		&11.81(4)	&11.22(7)\\
\bottomrule[0.5pt]
\end{tabular}
\caption{Parameters obtained from the fitting of Eq. (\ref{eq:eq6}) to the $1000\ \textup{Oe}$ field susceptibility data of titanate and stannate. The van Vleck $\chi_{\mathrm{vV}}$ contribution to the susceptibility is given in units of $\mathrm{erg\ Oe^{-2}\ mol^{-1}}$. The best fittings in the interval from 5 to 30 K are shown in the insets of Fig. \ref{fig:suscept_154Sm}\hyperref[fig:suscept_154Sm]{(a)} and \hyperref[fig:suscept_154Sm]{(b)}.}\label{tab:tab11}
\end{table}


The fittings were performed between two different intervals: $[5,20]\ \mathrm{K}$ and $[5,30]\ \mathrm{K}$. The results for both are shown in Table \ref{tab:tab11}. In the insets of Fig. \ref{fig:suscept_154Sm}, the best fitting performed in the interval $[5,30]\ \mathrm{K}$ is shown. Overall, for both intervals the parameters are in agreement with the titanate values reported by Singh \emph{et al.} \cite{PhysRevB.77.054408}, and the stannate values reported by Bondah-Jagalu \emph{et al.} \cite{nrc_cjp79_1381}. Nevertheless, the Curie-Weiss temperature of the stannate appears to be more sensitive to the increase of the upper-limit of the temperature interval. Paradoxically, $\theta_\mathrm{CW}$ changes from $0.11(6)\ \mathrm{K}$, signalising ferromagnetism, to $-0.44(7)\ \mathrm{K}$, or predominantly antiferromagnetic interactions. In Bondah-Jagalu \emph{et al.} \cite{nrc_cjp79_1381}, which performed the fittings within the interval $[5,20]\ \mathrm{K}$, a higher, ferromagnetic $\theta_\mathrm{CW}=1.36(21)\ \mathrm{K}$ is reported. Below, neutron diffraction results are going to demonstrate that the stannate effectively develops a long-range antiferromagnetic order. Clearly, however, we cannot conclude which interactions are predominant considering only the Curie-Weiss fitting.


\subsection{Long-range order}

\begin{figure*}
\includegraphics[trim={7.5cm 0cm 1.2cm 0cm},clip,width=6cm,angle=270]{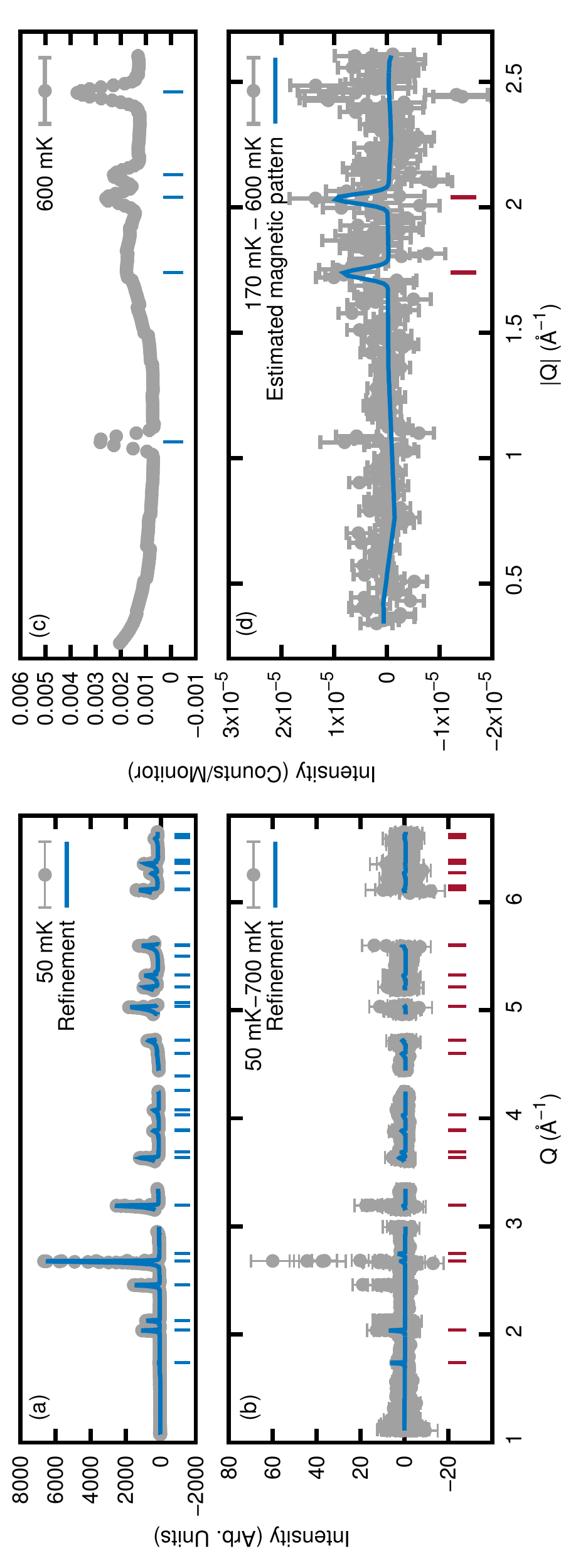}
\caption{Diffraction patterns of $\mathrm{^{154}Sm_2Ti_2O_7}$. (a) Data collected at the backscattering bank of WISH ($2\theta\sim152^{\circ}$) at a temperature of $700\ \mathrm{mK}$. The regions containing Bragg peaks from the copper sample can or from the insert aluminium are omitted. (b) Subtraction of data collected at 700 mK from data collected at 50 mK. (c) DNS data measured at 600 mK. The hump seen around $|\mathrm{Q}|=1.7\ \textup{\AA}^{-1}$ is background related to the deuterated alcohol used to promote cooling. (d) Subtraction of data collected at 600 mK from data collected at 170 mK. The continuous line in (a) and the tic marks in (a) and (c) show the refinement and position of nuclear Bragg peaks. The continuous line and the tic marks in (b) and (d) show the refinement and the position of the expected magnetic Bragg peaks of a structure corresponding to the irreducible representation $\Gamma_3$ (see text). } \label{fig:SmTO_mag_mon}
\end{figure*}

Fig. \ref{fig:SmTO_mag_mon}\hyperref[fig:SmTO_mag_mon]{(a)} displays data collected at the highest resolution detector bank of WISH (average $2\theta\sim152^{\circ}$) at $700\ \mathrm{mK}$ for $\mathrm{^{154}Sm_2Ti_2O_7}$. The diffraction pattern shown is interrupted around the peaks belonging to either the sample can (Cu) or to the dilution insert (Al) Bragg peaks. Only the momentum transfer range relevant to the observation of magnetism in Sm, based on the $F^2(|\mathbf Q|)$ of Fig. \ref{fig:Ei50meV_Sm}\hyperref[fig:Ei50meV_Sm]{(a)}, is displayed.

We remind the reader that, in the titanate, the anomaly in heat capacity takes place at $T^{Ti}_\mathrm{N}=350\ \mathrm{mK}$. In order to investigate the presence of magnetism in the sample, the data collected at $700\ \mathrm{mK}$ is subtracted from data measured at $50\ \mathrm{mK}$. The resulting difference is shown in Fig. \ref{fig:SmTO_mag_mon}\hyperref[fig:SmTO_mag_mon]{(b)}. Evidently, the pattern of Fig. \ref{fig:SmTO_mag_mon}\hyperref[fig:SmTO_mag_mon]{(b)} mirrors the intensities of the original peaks shown in Fig. \ref{fig:SmTO_mag_mon}\hyperref[fig:SmTO_mag_mon]{(a)}. The relevant differences measured above the background level appear all at structural Bragg peak positions. Moreover, the intensity of the difference peaks correlates strongly with the intensity of the nuclear reflections.

Fig. \ref{fig:SmTO_mag_mon}\hyperref[fig:SmTO_mag_mon]{(c)} shows the diffraction pattern measured at DNS for $\mathrm{^{154}Sm_2Ti_2O_7}$, at a temperature of 600 mK. Fig. \ref{fig:SmTO_mag_mon}\hyperref[fig:SmTO_mag_mon]{(d)} displays the difference between low (170 mK) and high temperature diffraction patterns. The correlation between the measured peak intensities in the high temperature and in the difference pattern is less obvious, since the weaker $220$ and $331$ reflections [red tics in Fig. \ref{fig:SmTO_mag_mon}\hyperref[fig:SmTO_mag_mon]{(b)} and \hyperref[fig:SmTO_mag_mon]{(d)}] have an intensity difference comparable with the stronger $111$ and $400$ structural Bragg peaks. Even so, analogously to the WISH data, the DNS measurement does not offer a clear-cut sign of sample magnetism.

In Mauws \emph{et al.} \cite{PhysRevB.98.100401}, polarised and unpolarised neutron diffraction suggested the presence of an all-in-all-out long-range order in a single crystal sample of $\mathrm{^{154}Sm_2Ti_2O_7}$. In their work, the intensity of the $220$ Bragg reflection is shown to present a sharp decrease at $T^{Ti}_\mathrm{N}$ upon warming from the base temperature. The ordered magnetic moment reported in Ref. \cite{PhysRevB.98.100401} is of $0.44(7)\mu_\mathrm{B}$, which is in agreement with the ground-state magnetic moment calculated with their crystal field analysis (see Table \ref{tab:tab10}). 

Comparing the nominal experimental conditions between our DNS data and the measurements of Ref. \cite{PhysRevB.98.100401}, it is possible that the temperature of 170 mK is not low enough for our sample magnetisation to reach full saturation. Nevertheless, in the sample of Mauws \emph{et al.}, as far as we can visually determine from the published data, the $220$ peak intensity already reaches $\sim90\%$ of its maximum just below $200\ \mathrm{mK}$. We believe that it is unlikely that the temperature of our powder did not go through $T^{Ti}_\mathrm{N}$ in any of the two experiments at WISH and DNS. Given that, again relying on the $220$ intensity data of Ref. \cite{PhysRevB.98.100401}, the onset of the ordered phase is very sharp, we expect that small temperature differences from $T^{Ti}_\mathrm{N}$ would produce visual effects on the measured diffraction pattern.

\begin{figure}[h]
\includegraphics[trim={0cm 0cm 0cm 0cm},clip,width=6cm,angle=270]{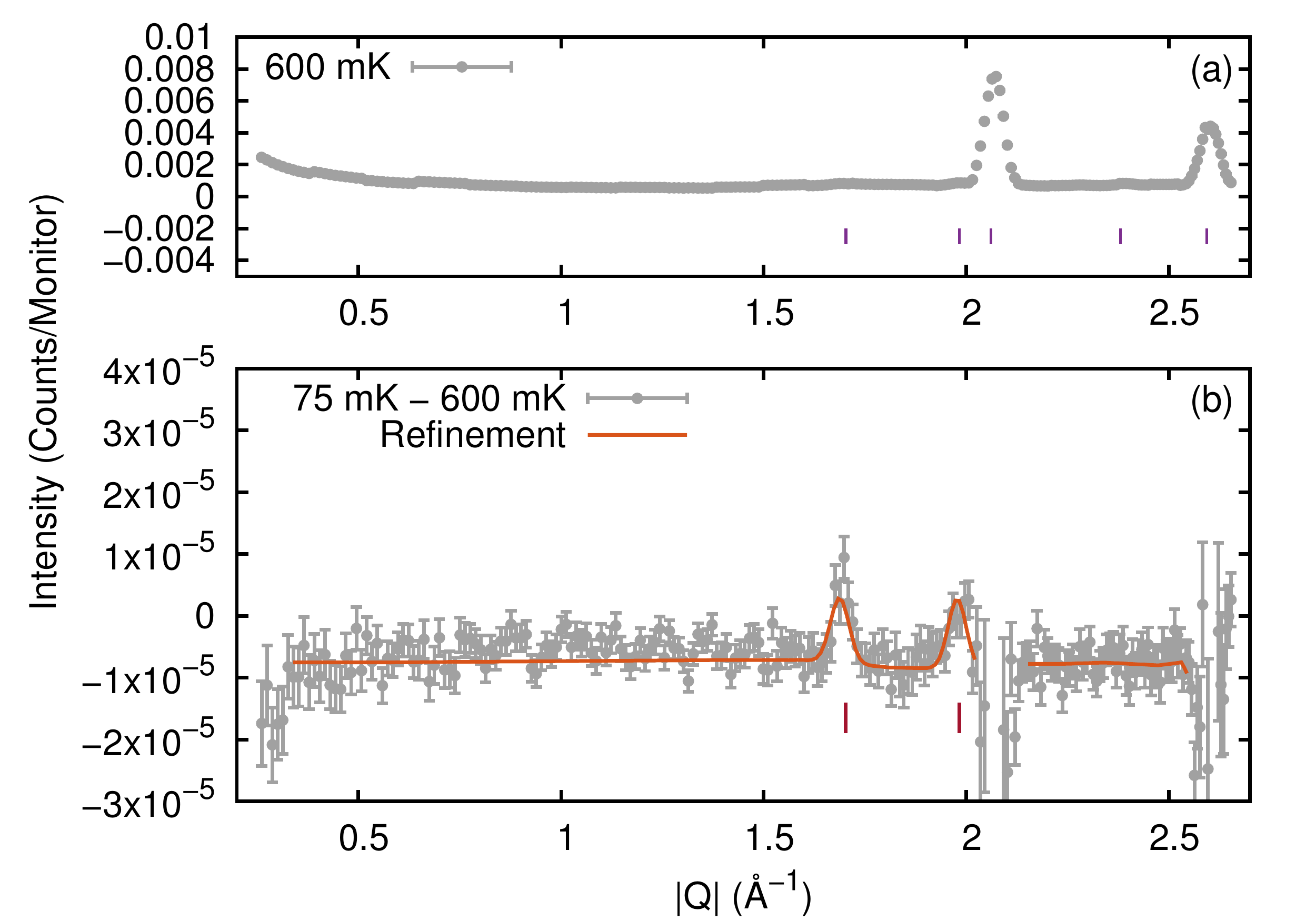}
\caption{Diffraction patterns of the $\mathrm{^{154}Sm_2Sn_2O_7}$ sample measured at DNS. (a) Data collected at a temperature of 600 mK is displayed. (b) Subtraction of data collected at 600 mK from data collected at 75 mK. The tic marks in (a) highlight the position of structural reflections, while in (b) the red tics mark the position of expected magnetic Bragg peaks for a structure corresponding to the irreducible representation $\Gamma_3$.}\label{fig:DNS_SmSO_mag_mon}
\end{figure}

\begin{table}
\tabcolsep=7pt
\renewcommand{\arraystretch}{1.3}
\begin{tabular}{c c c c c c c}
\toprule[1pt]
Site & X& Y & Z & m$^a$($\mu_\mathrm{B}$)& m$^b$($\mu_\mathrm{B}$) & m$^c$($\mu_\mathrm{B}$)\\
\cmidrule[0.3pt]{1-7}
1& $\frac{1}{2}$ & $\frac{1}{2}$ & $\frac{1}{2}$ & 0.14(1) & 0.14(1) & 0.14(1)\\
2& $\frac{1}{4}$ & $\frac{1}{4}$ & $\frac{1}{2}$ & -0.14(1) & -0.14(1) & 0.14(1)\\
3& $\frac{1}{2}$ & $\frac{1}{4}$ & $\frac{1}{4}$ & 0.14(1) & -0.14(1) & -0.14(1)\\
4& $\frac{1}{4}$ & $\frac{1}{2}$ & $\frac{1}{4}$ & -0.14(1) & 0.14(1) & -0.14(1)\\
\bottomrule[0.5pt]
\end{tabular}
\caption{Magnetic moments obtained in the refinement of $\mathrm{^{154}Sm_2Sn_2O_7}$ data of Fig. \ref{fig:DNS_SmSO_mag_mon}\hyperref[fig:DNS_SmSO_mag_mon]{(b)}. m$^a$, m$^b$ and m$^c$ are the magnetic moment components along the global $x, y\ \textup{and}\ z$ axis, respectively.}\label{tab6}
\end{table}

The data obtained at DNS for the stannate sample at 600 mK is shown in Fig. \ref{fig:DNS_SmSO_mag_mon}\hyperref[fig:DNS_SmSO_mag_mon]{(a)}, while the subtraction between low and high temperature is shown in Fig. \ref{fig:DNS_SmSO_mag_mon}\hyperref[fig:DNS_SmSO_mag_mon]{(b)}. As we have predicted, the study of one Sm-pyrochlore would greatly support the analysis performed on the other. Differently from the titanate, the diffraction pattern of $\mathrm{^{154}Sm_2Sn_2O_7}$ does display clear magnetic peaks in the difference plot. Only two of them are observed within the DNS momentum transfer limit, corresponding to the also structural $220$ and $311$ positions. Representation analysis performed in the $Fd\bar{3}m$ space group shows that the propagation vector $\mathbf{Q}=0$ magnetic structure that displays uniquely those reflections is the all-in-all-out arrangement, with irreducible representation $\Gamma_3$. The best refinement is shown along with the magnetic scattering data in Fig. \ref{fig:DNS_SmSO_mag_mon}\hyperref[fig:DNS_SmSO_mag_mon]{(b)}. The total refined magnetic moment $\mu_\mathrm{ref}=0.25(2)\ \mu_{\mathrm{B}}$ is in remarkable agreement with the one predicted in the crystal field analysis, namely, $\langle \hat\mu^{0}_z \rangle=0.27\ \mu_{\mathrm{B}}$. The individual components of the magnetic moment along each one of the global crystallographic axis is shown in Table \ref{tab6}, for one tetrahedron unit.

With the results for the stannate in hands, we return to the problem of the $\mathrm{^{154}Sm_2Ti_2O_7}$. The unambiguous long-range order cannot be shown, but a maximum value of ordered magnetic moment from WISH and DNS data, limited by background and noise level, can be definitely estimated. We repeat the representation analysis carried out on the stannate and determine a maximum $\mu_\mathrm{ref}\sim0.17\mu_{\mathrm{B}}$ from both datasets collected for the titanate. Those final fittings are superposed to the measured patterns in Fig. \ref{fig:SmTO_mag_mon}\hyperref[fig:SmTO_mag_mon]{(b)} and \hyperref[fig:SmTO_mag_mon]{(d)}.  

\section{Discussion}

Most of the magnetic entropy associated with the ground-state Kramers doublet of the Sm-based pyrochlores is recovered with the low temperature phase transition. Whereas some small difference to the estimated asymptotic $R\ \mathrm{ln(2)}$ is verified in Fig. \ref{HC_entropy_multiplot}\hyperref[HC_entropy_multiplot]{(b)}, hardly any significance can be attributed to it, given the complete disregard of other contributions, beyond electronic magnetism, to the heat capacity below $10\ \mathrm{K}$.

Given that, without high pressure synthesis, a stable, cubic pyrochlore phase can only be formed for rare-earths heavier than Sm$^{3+}$, usually the Stevens operator formalism can be unrestrictedly applied in the examination of the single-ion Hamiltonian of pyrochlores of the titanate family \cite{0953-8984-24-25-256003}. However, as different multiplets become closer in energy, the intermultiplet admixture of levels becomes more and more relevant in the ground-state wave function of the system \cite{OSBORN19911}. It was shown here that $\mathrm{Sm_2Ti_2O_7}$ and $\mathrm{Sm_2Sn_2O_7}$ illustrate this situation. 

The fitted $\mu_\mathrm{eff}$ of the samples are in very good agreement with the $\langle \hat\mu^{0}_z \rangle$ predicted by the CEF analysis. Despite the slightly smaller ordered moment, the magnitude of $\theta_\mathrm{CW}$ is higher in $\mathrm{^{154}Sm_2Ti_2O_7}$ than in $\mathrm{^{154}Sm_2Sn_2O_7}$ for the same temperature interval. Making a na\"ive association with our heat capacity results, it is possible that the stronger interactions in the titanate cause the tenuous suppression, relative to the stannate, observed in the phase transition temperature. 

Curiously, it was not possible to infer from the Curie-Weiss fittings if the stannate presents predominantly ferro- or antiferromagnetic correlations. This result is surprising, especially because a considerable change in the relative population of the CEF levels, which would alter the van Vleck susceptibility and the effective magnetic moment, is not expected to happen between 20 and 30 K. 

Neutron diffraction was used to show that $\mathrm{^{154}Sm_2Sn_2O_7}$ develops long-range dipolar all-in-all-out order below $T^{Sn}_\mathrm{N}=440\ \mathrm{mK}$. On the other hand, neither WISH or DNS data display consistent evidence of the presence of magnetic long-range order in $\mathrm{^{154}Sm_2Ti_2O_7}$. This result is somewhat surprising, but not completely unexpected. In view of the strong phase transition shown in Fig. \ref{HC_entropy_multiplot}, a magnetic long-range order should be definitely speculated. Meanwhile, we have demonstrated that the CEF ground-state magnetic moment of $0.16\mu_\mathrm{B}$ may be overly small to be detected by neutron diffraction in a powder sample. Even though thermal equilibration in this situation cannot be always guaranteed, our results are not subject to some accuracy limiting effects related to the estimation of magnetic moments in single crystal samples, such as, for example, absorption.

Independently of the dipolar static magnetism, there is still some room to a less conventional behaviour. The Sm-pyrochlores belong to the class of materials that possess dipolar-octupolar doublet ground-states \cite{PhysRevB.98.100401,PhysRevLett.115.197202}. In addition to the dipolar matrix elements connecting the $\ket{\pm0}$ states, which would result in the ordered magnetic moment calculated using Eq. (\ref{eq:eq5}), the next most important contribution to the total magnetic moment results from the matrix elements of the octupolar-operator \cite{PhysRevLett.115.197202} connecting the states $\ket{{^6}H_{9/2},\pm\frac{3}{2}}$ with $\ket{{^6}H_{9/2},\pm\frac{9}{2}}$. 
An interesting question to be addressed is whether the ground-state associated octupolar moments order or remain dynamic at low temperatures. It is possible that the system undergoes an octupolar-ordering, as predicted by the $XYZ$ model of Ref. \cite{PhysRevLett.112.167203}. Similarly to the dipolar case, the octupolar-ordering gives rise to a symmetry breaking resulting in a phase transition in heat capacity. Another possibility, already mentioned in Ref. \cite{PhysRevB.98.100401}, is that octupolar-octupolar coupling induces the magnetic-moment fragmentation, in a similar fashion to the studied in the pyrochlore $\mathrm{Nd_2Zr_2O_7}$ \cite{PhysRevLett.115.197202,petit}. 

Neutron diffraction probes directly only dipolar-ordering, while octupolar interactions can be indirectly inferred from the magnetic excitations present in the material \cite{PhysRevLett.115.197202,petit}. As the matrix elements of the octupolar tensor in $\mathrm{^{154}Sm_2Ti_2O_7}$ and $\mathrm{^{154}Sm_2Sn_2O_7}$ are significantly smaller that those in $\mathrm{Nd_2Zr_2O_7}$ \cite{PhysRevB.92.224430,PhysRevLett.115.197202}, possibly neutron scattering techniques are not, after all, ideal to check any of those in Sm-based pyrochlores. In this regard, more studies, especially measurements of dynamic susceptibility at low temperatures, would be indispensable.  

\section{Conclusion}

This work was dedicated to the study of the magnetic frustration manifested in two Sm-based pyrochlores, the titanate $\mathrm{^{154}Sm_2Ti_2O_7}$ and the stannate $\mathrm{^{154}Sm_2Sn_2O_7}$. It has been shown that both compounds undergo phase transitions at 350 and 440 mK, respectively. The magnetic entropy recovered with those phase transitions for $\mathrm{T}<8\ \mathrm{K}$ nearly reaches the expected from a ground-state doublet. We carried out inelastic neutron scattering measurements to investigate the crystal electric field excitations present in the compounds and find the single-ion, ground and excited-state wave functions that closely describe the measured static magnetic susceptibility of stannate and titanate, specially the unusual non-linear van Vleck contribution. 

At low temperatures, the development of long-range order in the samples below the phase transition temperature is investigated. The stannate ordered magnetic moment agrees with the dipolar-moment calculated in the crystal-field analysis. As for the titanate, if our data does not support the development of a dipolar long-range order below the phase transition temperature, it also cannot be used to demonstrate its absence for a $\mu_\mathrm{ref}<0.17\mu_{\mathrm{B}}$. It is still an open question whether the Sm-based pyrochlores present some exotic phase emerging from the characteristic ground-state dipolar-octupolar doublets. We demonstrate here, however, that both compounds present exciting prospects. 

\section{Acknowledgements}
The observation of the intermultiplet transitions in this work was only possible due to the indispensable advice of Dr. Jianhui Xu and Prof. Bella Lake. We also express our gratitude to Prof. Andrew Boothroyd, for his extremely generous support in the use of {{\normalsize S}{\footnotesize PECTRE}}, as well as discussions about the samarium particularities. Low temperature measurements were kindly assisted by the cryogenics teams of ISIS and FRM II (Andi, Helga and Heiner, herzlichen Dank). The work was supported by the Science and Technology Facilities Council STFC. V.P.A. was partially supported by CNPq-Brasil.
\bibliographystyle{apsrev4-1}
\bibliography{abragam,SUBRAMANIAN198355,S0924013699002113,S0925838817304541,S0955221914003628,kennedy_sclengths,neutron_sclengths,Carnall,Hutchings,wybourne,Malkin,PhysRevB.77.054408,PhysRevB.93.214308,PhysRevB.98.100401,spectre,FREEMAN197911,vanvleck,xu_jianhui,RevModPhys.82.53,PhysRevLett.74.3423,PhysRevB.54.4276,osborn_book,Bramwell_suscept,PhysRevB.92.224430,rosenkranz,vandeborre,merlin,wish,PhysRevLett.112.167203,PhysRevLett.108.037202,petit,PhysRevLett.115.197202,stevens,nrc_cjp79_1381,Balents,0953-8984-24-25-256003,Blundell,OSBORN19911,PhysRevB.91.224430}

\end{document}